\documentclass[11pt]{article}
\pdfoutput=1

\usepackage{jcappub}
\usepackage{graphicx} 	
\usepackage[justification=justified, singlelinecheck=off, compatibility=false]{caption}
\usepackage{amsmath}	
\usepackage{bm}		
\usepackage{amssymb}	
\usepackage{color} 		
\usepackage{times} 		
\usepackage{tensor} 	
\usepackage{float}
\usepackage{hyperref} 
\hypersetup{
    colorlinks=true,       
    linkcolor=red,          
    citecolor=blue,        
    filecolor=magenta,      
    urlcolor=blue           
}

\usepackage{ulem}


\numberwithin{equation}{section}

\usepackage{myterms}

\newcommand{\AB}{\text{\sc ab}}

\newcommand{\gX}{g_{\text{\sc x}}}

\newcommand{\kup}{{\rm k}}

\newcommand{\Pcal}{\mathcal{P}}
\newcommand{\Xbb}{\mathbb{X}}

\newcommand{\gHstr}{g^{{\text{\sc h}}}_{\rm str}}
\newcommand{\gbarHstr}{\bar{g}^{{\text{\sc h}}}_{\rm str}}
\newcommand{\gHHstr}{g^{\text{\sc hh}}_{\rm str}}
\newcommand{\gZstr}{\bar{g}^{{\text{\sc z}}}_{\rm str}}
\newcommand{\gZZstr}{\bar{g}^{{\text{\sc zz}}}_{\rm str}}

\newcommand{\GHHcusp}{\Gamma_{\text{\sc hh}}^{\rm (cusp)}}

\newcommand{\GHHkk}{\Gamma_{\text{\sc hh}}^{\rm (k-k)}}

\newcommand{\GHcusp}{\Gamma_{\text{\sc h}}^{\rm (cusp)}}

\newcommand{\GHkk}{\Gamma_{\text{\sc h}}^{\rm (k-k)}}

\newcommand{\Gcusp}{\Gamma^{\rm (cusp)}}

\newcommand{\Gkk}{\Gamma^{\rm (k-k)}}

\newcommand{\idx}{p}
\newcommand{\pp}{\Gamma}

\newcommand{\Pcusp}{P^{\rm (cusp)}}

\newcommand{\Pkk}{P^{\rm (k-k)}}

\newcommand{\PHHcusp}{P_{\text{\sc hh}}^{\rm (cusp)}}

\newcommand{\PHHkk}{P_{\text{\sc hh}}^{\rm (k-k)}}

\newcommand{\FIRAS}{\text{\sc firas}}
\newcommand{\PIXIE}{\text{\sc pixie}}
\newcommand{\BBN}{\text{\sc bbn}}

\newcommand{\IceCube}{{\rm IceCube}}

\newcommand{\CR}{\text{\sc cr}}

\newcommand{\mudist}{\mu_{\rm dist.}}

\newcommand{\Lcuspy}{L_{\rm cuspy}}
\newcommand{\tcuspy}{t_{\rm cuspy}}
\newcommand{\tdecay}{\tau}
\newcommand{\Lkinky}{L_{\rm kinky}}
\newcommand{\tkinky}{t_{\rm kinky}}
\newcommand{\burst}{\rm burst}

\newcommand{\Hcond}{\eta_{0}}

\newcommand{\oef}{0.2}
\newcommand{\tss}{0.3}
\newcommand{\tff}{0.2}
\newcommand{\oet}{0.2}
\newcommand{\zest}{0.1}
\newcommand{\zsss}{0.1}
\newcommand{\zffn}{0.06}
\newcommand{\zons}{0.02}
\newcommand{\fos}{0.4}
\newcommand{\zoff}{0.02}
\newcommand{\sff}{0.7}
\newcommand{\ses}{0.8}
\newcommand{\tsn}{3}
\newcommand{\foo}{0.4}
\newcommand{\stf}{0.7}
\newcommand{\tfn}{0.3}
\newcommand{\tto}{2}
\newcommand{\ost}{0.2}

\begin{document}

\title{Cosmic Strings in Hidden Sectors:  2.  Cosmological and Astrophysical Signatures}

\date{\today}

\author{Andrew J. Long}
\emailAdd{andrewjlong@asu.edu}
\author{and Tanmay Vachaspati}
\emailAdd{tvachasp@asu.edu}
\affiliation{Physics Department, Arizona State University, Tempe, Arizona 85287, USA.}

\abstract{

Cosmic strings can arise in hidden sector models with a spontaneously broken Abelian symmetry group.  
We have studied the couplings of the Standard Model fields to these so-called dark strings in the companion paper.  
Here we survey the cosmological and astrophysical observables that could be associated with the presence of dark strings in our universe with an emphasis on low-scale models, perhaps TeV.  
Specifically, we consider constraints from nucleosynthesis and CMB spectral distortions, and we calculate the predicted fluxes of diffuse gamma ray cascade photons and cosmic rays.  
For strings as light as ${\rm TeV}$, we find that the predicted level of these signatures is well below the sensitivity of the current experiments, and therefore low scale cosmic strings in hidden sectors remain unconstrained.  
Heavier strings with a mass scale in the range $10^{13} \GeV$ to $10^{15} \GeV$ are at tension with nucleosynthesis constraints.  
}

\maketitle

\section{Introduction}\label{sec:Introduction}

A hidden sector is a well-motivated and phenomenologically rich extension of the Standard Model (SM) with interesting implications for dark matter \cite{ArkaniHamed:2008qp, Cassel:2009pu, Chun:2010ve, Chu:2011be, Baek:2014jga, Basak:2014sza} and collider physics \cite{Leike:1998wr, Langacker:2008yv, Djouadi:2012zc, Alves:2013tqa, No:2013wsa}.  
In one minimal construction, the SM is extended to include a complex scalar field $S$ charged under a $\U{1}$ gauge group with $\hat{X}_{\mu}$ as its vector potential.  
The only coupling to the SM is through the Higgs field $\Phi$ and hypercharge gauge field $Y_{\mu}$, and the interaction is written as 
\begin{equation}\label{eq:Lint}
	\Lcal_{\rm int} = -\alpha \, \Phi^\dag \Phi S^{*} S - \frac{\sin\epsilon}{2} {\hat X}_{\mu\nu}Y^{\mu\nu}
\end{equation}
where $\alpha$ is the Higgs portal coupling and $\sin \epsilon$ is the gauge kinetic mixing parameter.  
If the mass scale in the hidden sector is $M \lesssim \TeV$, then $\alpha$ and $\sin \epsilon$ are constrained by various laboratory tests \cite{Jaeckel:2013ija,Belanger:2013kya}, but these parameters are as yet unconstrained if the scale of new physics is above TeV.  

If the the $\U{1}$ of the hidden sector is spontaneously broken, then the model admits cosmic string solutions \cite{VilenkinShellard:1994} know as dark strings \cite{Vachaspati:2009jx}.  
If the hidden sector is low scale, perhaps $M \sim 10 \TeV$, 
then the standard gravitational probes of cosmic strings are ineffectual, and one must turn to the particle physics interactions.  
In principle, the cosmic strings could provide an indirect probe of interactions in \eref{eq:Lint} even if the scale of the hidden sector is well above TeV.  
This is because the strings would persist as relics today, and do not have to be produced in the lab.  
In this sense, these astrophysical probes of cosmic strings are the same as those examined in the indirect detection of dark matter \cite{Porter:2011nv}.  

Although dark strings result from topology in the hidden sector, we found in \rref{Hyde:2013fia} that they are dressed with condensates of SM fields, namely the Higgs field (previously noted by \rref{Peter:1992}) and the Z field.  
The couplings in \eref{eq:Lint} and the dressing lead to effective interactions between the dark string, at spacetime location ${\mathbb X}^\mu$, and the SM fields that are of the form \cite{Hyde:2013fia}\footnote{We use a slightly different notation here than Refs.~\cite{Hyde:2013fia, Long:2014mxa}, and hence the ``barred'' coupling constant.}
\begin{align}\label{eq:S_eff}
	S_{\rm eff} = \ & 
	\gHHstr  \, \int d^2\sigma \sqrt{-\gamma} \ \phi_H^2(\Xbb)
	+ \gbarHstr \, M \, \int d^2\sigma \sqrt{-\gamma} \ \phi_H({\mathbb X}) \nn
	& + \gZZstr \int d^2\sigma \, \sqrt{-\gamma} \, Z_{\mu}(\Xbb) Z^{\mu}(\Xbb) 
	+ \frac{\gZstr}{2} \int d\sigma^{\mu\nu} Z_{\mu\nu}({\mathbb X}) 
	\per
\end{align}
The quadratic Higgs interaction (first term) arises from the Higgs portal operator of \eref{eq:Lint}, and so $\gHHstr \approx \alpha$.  
The linear Higgs interaction (second term) arises from the Higgs condensate on the string, $\langle \Phi^0 \rangle = \Hcond$, and so $\gbarHstr \propto (\Hcond / M)$.  
The quadratic and linear Z-bosons interactions (third and fourth terms) arise from the mixings in the scalar and gauge sectors, and so $\gZZstr \sim (\gZstr)^2 \sim (\Hcond / M)^4$.  
The interactions in $S_{\rm eff}$ allow the dark strings to radiate Higgs and Z particles, which we studied in \rref{Long:2014mxa} (see also \sref{sec:SM_Particle_Prod} and references therein).

In the present paper we derive cosmological and astrophysical signatures of dark strings that arise from radiation of SM particles.  
In the literature, there has been extensive work on non-gravitational probes of cosmic string networks (references provided in \sref{sec:Observables}).  
Unlike the universal gravitational constraints, the probes that rely on particle emission from the string are model-dependent.  
For instance, the particle emission rate depends on parameters such as (i) the form of the coupling as in \eref{eq:S_eff}, (ii) whether or not the string is superconducting, and (iii) the configuration of the string that is radiating, {\it e.g.}, is it cuspy / kinky? is it large / small compared to the de Broglie wavelength of the radiated particle?
As such, the predicted observables available in the literature do not necessarily carry over to the dark string model.  
The present analysis builds upon and extends prior work in the following ways.  
\begin{enumerate}
	\item  We use the particle radiation rates calculated recently in \rref{Long:2014mxa}, which rectified a handful of errors in the literature.  
	\item  We pay special attention to light strings, which are motivated by ${\rm TeV}$ scale extensions of the SM.  The general consensus in the literature is that such light strings are unconstrained\footnote{Refs.~\cite{MacGibbon:1989kk, Vachaspati:2009kq} do find lower bounds on the string tension, but these calculations overlook non-gravitational radiation in the loop decay calculation, as pointed out by \cite{Bhattacharjee:1989vu}.  }, and our conclusions do not differ. In the course of the calculation, however, we uncover a previously overlooked suppression in the abundance of small string loops that decay non-gravitationally.  
	\item  We generalize previous calculations by including both populations of kinky and cuspy loops together.  
\end{enumerate}

The paper is organized as follows.  
We start in \sref{sec:SM_Particle_Prod} with a recap of previous results obtained in Refs.~\cite{Hyde:2013fia, Long:2014mxa}. 
The interactions of dark strings with SM particles affects their dynamics and also the properties of the dark string network {\it e.g.} the number distribution of loops.  
In \sref{sec:Friction_LoopDist_NetworkEvol} we build the cosmological scenario of dark strings, and provide expressions for the number density of closed loops and the Higgs radiation we expect from them.  
\sref{sec:Observables} converts the flux of Higgs radiation into observable signatures and compares predictions with current observations.  
The essential idea is that strings inject energy mainly in the form of Higgs particles into the cosmological medium, the Higgses then decay to photons and other particles which can be observed by experiments. 
In \sref{sec:Constraints} we apply the translate the constraints onto the underlying Lagrangian parameters.  
We summarize our findings in \sref{sec:Conclusion}.  
We calculate the loop length distribution in \aref{app:loop_dist}.  

Symbols that appear frequently are defined as follows:  $\mu = M^2$ is the string tension, $M$ is the scale of symmetry breaking in the hidden sector, $L$ is the length of a string loop, and $G \approx (1.22 \times 10^{19} \GeV)^{-2}$ is Newton's constant.  
Radiation-matter equality occurs at time $t_{eq} \approx 97\, 000 \yr$, redshift $z_{eq} \approx 2800$, and temperature $T_{eq} \approx 0.66 \eV$.  

\section{Radiation and Scattering from the Dark String}\label{sec:SM_Particle_Prod}

Cosmic strings may radiate both gravity waves and particles.  
Gravitational radiation is universal since the cosmic string's stress-energy tensor sources the gravitational field, but particle radiation is model-dependent since it requires a coupling of the radiated fields to the string-forming fields.  
Here we discuss the particle radiation from the dark string, whose couplings to the SM fields are known.  

Radiation requires an accelerated (or curved) segment of string.  
As such, radiation is most efficient from cusps and kinks on string loops where the curvature is high  \cite{Damour:2001bk}.  
A cusp is a point on the string where the local velocity momentarily approaches the speed of light \cite{Turok:1984cn}.  
Roughly $30-50\%$ of stable string loops are expected to possess cusps \cite{Copi:2010jw}.  
A kink forms when two string segments pass through one another and reconnect such that the tangent vector to the sting is discontinuous.  
Since string loops form from such reconnections, all string loops are expected to possess kinks \cite{Garfinkle:1987yw}.  
Radiation is emitted from individual kinks and well as the collisions of two kinks \cite{Garfinkle:1987yw}.  

The couplings of the dark string \cite{Hyde:2013fia} allow it to radiate gravity waves, SM Higgs bosons, Z-bosons, and SM fermions \cite{Long:2014mxa}.  
We quantify the radiation into each channel using a power $P(t,L)$, which is the rate of energy loss of a loop of length $L$ at time $t$, {\it i.e.} the integral of the radiation spectrum averaged over one loop oscillation period $\Delta t = L / 2$.  
Provided that the loop is oscillating periodically (transient behavior has died away and backreaction is neglected), the power is independent of $t$.  
In general, the power depends on the shape of the loop, not just its length.  
However, the shape dependence does not affect the parametric relationship between $P$ and $L$ at leading order, but it simply affects the overall prefactor (see below).  

The rate of energy loss into gravity waves from a cusp or a kink is given by the standard expression \cite{Turok:1984cn, Vachaspati:1984gt} 
\begin{align}\label{eq:Pgrav}
	P_{\rm grav} = \Gamma_g \, G M^4
\end{align}
where $\mu = M^2$ is the string tension.  
The dimensionless coefficient $\Gamma_g$ encodes our ignorance of the shape of the loop, and for typical loops it is $50 \lesssim \Gamma_g \lesssim 100$.  

Particle emission from cosmic strings has been studied extensively, and the calculation has been revisited intermittently over the years \cite{Srednicki:1986xg, Brandenberger:1986vj, Damour:1996pv, Peloso:2002rx, Vachaspati:2009kq, Berezinsky:2011cp, Dufaux:2012np, Lunardini:2012ct, Long:2014mxa}.  
This continuing interest is most likely because the radiation spectrum differs from model to model depending on the form of the coupling and additionally, unlike gravitational radiation, the particle radiation power is different for cusp, kinks, and kink-kink collisions.  
\rref{Long:2014mxa} found that when fields are coupled to the string as in \eref{eq:S_eff}, it is generally true that the radiation from cusps and kink-kink collisions has an average power that scales with loop length $L$ as 
\begin{align}\label{eq:P_sim_L}
	\Pcusp \sim 1 / \sqrt{L} 
	\qquad {\rm and} \qquad
	\Pkk \sim 1 / L \per  
\end{align}
Additionally, the radiation from individual kinks is only nonzero for small loops $L < M^2 / m^3$ with $m$ the mass of the particle being radiated.  
This result reconciled a number of conflicting claims in the literature, and it holds true for any light field coupled to a cosmic string as in \eref{eq:S_eff}.  

The magnitudes of the dimensionful coefficients in \eref{eq:P_sim_L}, however, are not so universal.  
In the limit where the radiated particle's mass is small compared to the string scale, $m \ll M$, one might naively estimate the power using dimensional analysis to be $\Pcusp \approx g^2 M^2 / \sqrt{ML}$ where $g^2 \leq O(1)$ is a coefficient that depends on the coupling constants.  
A careful calculation reveals that $\Pcusp$ can be larger by as much as $\sqrt{M/m}$, or it can be smaller by as much as $m^2 / M^2$.  
As explained in \rref{Long:2014mxa}, the failure of the naive calculation is in the hierarchal nature of the problem:  $L^{-1} \ll m \ll M$.  

In order to keep our calculation general at the outset, we will parametrize the average radiation power by cusps and kink collisions as
\begin{align}
	\Pcusp &= \Gcusp \frac{M^2}{\sqrt{m L}} \label{eq:P_cusp} \\
	\Pkk &= \Gkk \frac{M^2}{m L} \label{eq:P_kk} 
\end{align}
where $\Gcusp$ and $\Gkk$ are dimensionless coefficients.  
We will focus on Higgs boson radiation for which the mass of the particle being radiated is $m = m_H \approx 125 \GeV$.  
Since $M$ is the largest mass scale in the problem and $m$ is the smallest mass scale, we expect $\Gcusp, \Gkk < O(1)$ in general. 
However, it is important to emphasize that $\Gcusp$ and $\Gkk$ may themselves scale like some power of $(m/M)$ (see below), and thus, the explicit dependence on $M$ and $m$ shown in \erefs{eq:P_cusp}{eq:P_kk} is not always indicative of the actual parametric dependence.  

The model-dependence is contained within the coefficients $\Gcusp$ and $\Gkk$.  
The radiation of $Z$-bosons is typically subdominant owing to the factors of $\gZstr \sim (\Hcond/M)^2$ and $\gZZstr \sim (\Hcond/M)^4$ in $S_{\rm eff}$, \eref{eq:S_eff}.  
For Higgs radiation arising from the quadratic coupling in $S_{\rm eff}$ we have \cite{Long:2014mxa}
\begin{align}\label{eq:Gamma_HH}
	\GHHcusp = (10^{-5} - 10^{-2}) (\gHHstr)^2 \sqrt{\frac{m}{M}}
	\qquad {\rm and} \qquad
	\GHHkk = (10^{-4} - 10^{-1}) (\gHHstr)^2 \frac{m}{M}
\end{align}
where the ranges quantify our ignorance of the string loop shape, and $\gHHstr \approx \alpha$ is the Higgs portal coupling.  
For Higgs radiation arising from the linear coupling in $S_{\rm eff}$ we have instead\footnote{In this case, $\Pcusp$ and $\Pkk$ diverge as $m \to 0$, but the derivation of \erefs{eq:P_cusp}{eq:P_kk} breaks down for $mL<1$.  Although this regime is not relevant to the case of Higgs boson radiation, we note that the powers will go as $\Pcusp \sim \Pkk \sim \Hcond^2$ \cite{Damour:1996pv}.  }   
\begin{align}\label{eq:Gamma_H}
	\GHcusp = (10^{-4} - 10^{-1}) (\gHstr)^2 \frac{\Hcond^2}{M^2}
	\qquad {\rm and} \qquad
	\GHkk = (10^{-2} - 10^{1}) (\gHstr)^2 \frac{\Hcond^2}{M^2} 
\end{align}
where we have written $\gbarHstr = \gHstr (\Hcond / M)$, and \rref{Hyde:2013fia} found $\gHstr \approx 10 \alpha / \kappa$ with $\kappa$ the self-coupling of the string-forming scalar $S(x)$.  
There is unfortunately no general analytic expression for the value of the Higgs condensate at the string core, $\Hcond$, in terms of the model parameters.  
\rref{Hyde:2013fia} numerically investigated a region of parameter space with $\kappa \approx \lambda \approx 10 \alpha \approx 0.1$ finding $\Hcond \approx \eta$; see their Fig 5b.  
In this regime, the Higgs radiation via quadratic coupling, \eref{eq:Gamma_HH}, dominates over the Higgs radiation via linear coupling, \eref{eq:Gamma_H}.  
Building on earlier work \cite{Haws:1988ax}, \rref{Mota:2014uka} recently argued that the parameter regime $\kappa \ll \alpha < \sqrt{\lambda \kappa}$ yields a solution with a large condensate, $\Hcond \approx \sqrt{\alpha / \lambda} \, M$.  
In this regime, the linear coupling dominates over the quadratic coupling.  
Additionally, \rref{Mota:2014uka} estimated the Higgs radiation power due to non-perturbative cusp evaporation \cite{Brandenberger:1986vj}.  
They found the power takes the same form as $\Pcusp$ in \eref{eq:P_cusp} when $\Gcusp$ is identified as a cusp evaporation ``efficiency factor,'' but this parameter is treated as free variable and not expressed in terms of the couplings of the underlying Lagrangian.  
For the time being, we will use \erefs{eq:P_cusp}{eq:P_kk} assuming only that $\Gcusp, \Gkk < 1$.  
Then we will return to the issue of model dependence in \sref{sec:Constraints}.  

In deriving the particle radiation powers represented by \erefs{eq:Gamma_HH}{eq:Gamma_H}, it was assumed in \rref{Long:2014mxa} that the string can be approximated using the zero thickness Nambu-Goto string model.  
It has been argued by Refs.~\cite{Vincent:1997cx, Hindmarsh:2008dw} that this approximation does not capture a non-perturbative radiation channel, which can be seen when using the finite thickness Abelian-Higgs string model.  
However, the Abelian-Higgs results have been criticized, {\it e.g.} by Refs.~\cite{Moore:1998gp, Olum:1999sg} where it is claimed that the Abelian-Higgs string network simulations lack dynamic range to reliably assess the particle production.  
Here we choose to focus on the Nambu-Goto case; for a discussion of constraints in the Abelian-Higgs case, see \rref{Mota:2014uka}.  

In \rref{Long:2014mxa} we also studied the scattering of SM fermions from the dark string.  
Provided that the gauge kinetic mixing parameter $\sin \epsilon$ is not too small,
the scattering proceeds primarily through the Aharonov-Bohm (AB) interaction \cite{Aharonov:1959fk}.  
The transport cross section takes the form \cite{Alford:1988sj}
\begin{align}\label{eq:sigt_AB}
	\sigma_t({\bf k}) = \frac{2}{\kup_{\perp}} \sin^2 (\pi \phi_{\AB}) 
\end{align}
where $\kup_{\perp}$ is the magnitude of the momentum in the plane transverse to the string and $2 \pi \phi_{\AB}$ is the phase that a particle acquires upon circling the string.  
For the dark string, the AB phases of the SM fermions were found to be \cite{Hyde:2013fia}
\begin{align}\label{eq:AB_phase}
	\phi_{\AB} = q \, \Theta
	\qquad {\rm with} \qquad
	\Theta = -2 \frac{\cos \theta_w \, \sin \epsilon}{\gX} 
\end{align}
where $q$ is the electromagnetic charge, $\gX$ is the gauge coupling of the new $\U{1}$ force, and $\theta_w$ is the weak mixing angle.  

\section{Evolution of the Dark String Network}\label{sec:Friction_LoopDist_NetworkEvol}

In this section we discuss the evolution of the string network from formation until today, and we calculate the properties of the network that are relevant for the calculation of observables in the following section.  

\subsection{Friction}\label{sub:Friction}

The evolution of a cosmic string network is typically friction dominated at formation due to the large elastic scattering cross section between the strings and the ambient plasma \cite{Vilenkin:1981kz}.  
If a string moves at velocity ${\bf v}$ through a plasma of temperature $T$ then elastic scatterings induce a drag force per unit length ${\bf f}_{\rm drag}$, which takes the form \cite{Vilenkin:1991zk}
\begin{align}\label{eq:fdrag_def}
	{\bf f}_{\rm drag} = - \beta T^3 \gamma {\bf v}
\end{align}
in the rest frame of the string.  
Here $\beta$ is the dimensionless drag coefficient and $\gamma = ( 1 - \abs{\bf v}^2 )^{-1/2}$ is the local Lorentz factor.  
The drag force smooths out features on the string, such as cusps and kinks, which prohibits radiation during the friction dominated era.  
In the presence of a friction force and an expanding background with Hubble parameter $H$, the string experiences an effective drag of $2 H + \beta T^3 / M^2$.  
Eventually the friction force becomes negligible below a temperature of \cite{Vilenkin:1981kz, Vilenkin:1991zk}
\begin{align}\label{eq:Tast_estimate}
	T_{\ast} \approx \frac{2 M^2}{\beta M_0} \approx 1 \eV \left( \frac{M}{10 \TeV} \right)^2 \left( \frac{\beta}{0.1} \right)^{-1} 
\end{align}
where $M_0 = T^2 / H \approx 10^{18} \GeV$ during the radiation dominated epoch.  
For a light string network with a sizable drag coefficient, the friction dominated phase can last until recombination ($T_{rec} \approx 0.1 \eV$).  
If the drag coefficient is small, however, friction domination may terminate much earlier.  

To determine if $\beta$ is large or small, we calculate it explicitly for the dark string interacting with the SM plasma.  
The drag coefficient is given by \cite{Vilenkin:1991zk}
\begin{align}\label{eq:beta_bound}
	\beta = \sum_{a} \frac{n_a(T)}{T^3} \sin^2 (\pi \phi_{\AB \, a}) \com
\end{align}
where the sum is over species in the plasma, $n_a(T)$ is the abundance of species $a$, and $\phi_{\AB \, a}$ is the corresponding AB phase.  
If a species is thermalized and relativistic, then $n_a(T) \approx T^3$ and $\beta \approx \sin^2 (\pi \phi_{\AB \, i})$.  
Although $\phi_{\AB \, a}$ is model-dependent, see \eref{eq:AB_phase}, it is not unreasonable to expect $\beta = O(0.1)$ as in the estimate of \eref{eq:Tast_estimate}.  
However, as species go out of equilibrium $n_a$ drops below $T^3$, and there is a corresponding decrease in $\beta$.  

Let us estimate $\beta$ before and after the era of electron-positron annihilation, which occurred at $T_{ann} \approx 0.1 \MeV$.  
For $T > T_{ann}$ the dominant contribution to $\beta$ comes from electrons and positrons, which are relativistic and have an abundance $n_{e^{\pm}}(T > T_{ann}) \approx T^3$.  
Then \eref{eq:beta_bound} gives the drag coefficient to be 
\begin{align}
	\beta(T > T_{ann}) 
	\approx \frac{n_{e^-}(T) + n_{e^+}(T)}{T^3} \sin^2 ( \phi_{\AB , e} ) 
	\approx  \sin^2 ( \phi_{\AB , e} ) \per
\end{align}
Assuming that annihilation is totally efficient, then for $T < T_{ann}$ the positron abundance is $n_{e^+} \approx 0$ and the electron abundance is $n_{e^-} \approx \eta_B T^3$ where $\eta_B \approx 10^{-10}$ is the baryon asymmetry of the universe.  
At this time, 
\begin{align}
	\beta(T < T_{ann}) 
	\approx 10^{-10} \sin^2 ( \phi_{\AB , e} ) \per
\end{align}
With such a small drag coefficient, one can verify that $| {\bf f}_{\rm drag} |$ is much less than the Hubble drag, for all values of the string mass scale in the range of interest, $M > \TeV$.  
Therefore we conclude that the dark string network remains friction dominated (by virtue of Aharonov-Bohm scattering) until such a time that the temperature is equal to 
\begin{align}\label{eq:Tast}
	T_{\ast} = {\rm Max} \Bigl[ \frac{M}{M_0 \, \sin^2 (\pi \phi_{\AB \, e})} \, , \, T_{ann} \Bigr]
\end{align}
where $M_0 \approx 10^{18} \GeV$.  
That is, friction domination must end at or before the era of electron-positron annihilation ($T_{ann} \approx 0.1 \MeV$).  

\subsection{Loop Decay}\label{sub:LoopDecay}

After the friction force becomes negligible, the string network begins to evolve freely.  
Long strings reconnect to form loops, and these loops shrink as they lose energy in the form of gravity waves and particle emission.  
A loop with energy $E$ (center of mass frame) has a length $L = E / M^2$ that evolves subject to
the loop decay equation 
\begin{align}\label{eq:loop_decay_eqn}
	M^2 \frac{dL}{dt} = - P(t,L)
\end{align}
where $P(t,L)$ is the average rate of energy loss from a loop of length $L$ at time $t$.  
This radiation arises dominantly from cusps and kinks, as we discussed in \sref{sec:SM_Particle_Prod}, and where all loops have kinks but only an $O(1)$ fraction are expected to possess cusps.  
It is convenient, then, to consider two populations of loops in the system: those that have cusps and kinks (cuspy loops) and those that have only kinks (kinky loops).  

\qquad \\
{\bf (i) Cuspy Loops} \\
For the case of cuspy loops, the loop decay equation becomes \cite{Vachaspati:2009kq} \begin{align}\label{eq:dLdt_cuspy}
	M^2 \frac{dL}{dt} = - \Gamma_{g} G M^4 - \Gcusp \frac{M^2}{(mL)^{1/2}} 
	\qquad , \qquad
	L(t_{i}) = L_{i}
\end{align}
where the two terms on the right-hand side are the rates of energy loss into gravity waves and particles, see \erefs{eq:Pgrav}{eq:P_cusp}.  
The kinks on these cuspy loops also radiate, see \eref{eq:P_kk}, but their contribution to the total power is suppressed since $ML \gg 1$.  
This system has the characteristic length and time scales \cite{Vachaspati:2009kq} 
\begin{align}
	\Lcuspy & \equiv \frac{[\Gcusp]^2}{(\Gamma_g G)^2} \frac{1}{M^4 m} 
	\label{eq:L_cuspy} \\
	\tcuspy & \equiv \frac{ [ \Gcusp]^2 }{(\Gamma_g G)^3} \frac{1}{M^6 m} 
	\label{eq:t_cuspy} \per
\end{align}
Comparing the terms in \eref{eq:dLdt_cuspy}, one can see that gravitational radiation is dominant for large loops ($L > \Lcuspy$) and particle radiation is dominant for small loops ($L < \Lcuspy$) \cite{Vachaspati:2009kq, Bhattacharjee:1989vu}.  
For the well-studied GUT-scale string, $M$ is sufficiently big that almost all sub-horizon scale loops today are considered large, and the particle radiation can be neglected.  
For the models that we are interested in, however, $M$ may be as low as ${\rm TeV}$, and in this regime particle radiation can be the dominant mode of energy loss.  

Integrating \eref{eq:dLdt_cuspy} gives 
\begin{align}\label{eq:yeqn_cuspy}
	y - 2 \sqrt{y} + 2 \ln ( 1 + \sqrt{y} ) = y_{i} - 2 \sqrt{y_{i}} + 2 \ln ( 1 + \sqrt{y_{i}} ) - (x - x_{i}) 
\end{align}
where $y = L / \Lcuspy$ and $x = t / \tcuspy$ are dimensionless length and time coordinates.  
An exact solution of \eref{eq:yeqn_cuspy} is not available, but it can be solved in the three limiting cases:  
\begin{align}\label{eq:length_cases}
\begin{array}{lclcl}
	\text{Case Ia:} & \quad & \text{large loops that were large at formation} & \quad & (\Lcuspy \ll L \leq L_i) \\
	\text{Case Ib:} & \quad & \text{small loops that were large at formation} & \quad & (L \ll \Lcuspy \ll L_i) \\
	\text{Case II:} & \quad & \text{small loops that were small at formation} & \quad & (L \leq L_i \ll \Lcuspy)
	\end{array}
\end{align}
The solutions are 
\begin{align}\label{eq:ysol_cuspy_1}
	L \approx \begin{cases}
	L_{i} - \frac{\Lcuspy}{\tcuspy} (t - t_{i}) 
	& \Lcuspy \ll L < L_{i} \\
	\bigl( \frac{3}{2} \bigr)^{2/3} \bigl[ L_{i} \, \Lcuspy^{1/2}- \frac{\Lcuspy^{3/2}}{\tcuspy} ( t - t_{i} ) \bigr]^{2/3}
	& L \ll \Lcuspy \ll L_{i} \\
	\bigl[ L_{i}^{3/2} - \frac{3}{2} \frac{\Lcuspy^{3/2}}{\tcuspy} (t - t_{i}) \bigr]^{2/3} 
	& L < L_{i} \ll \Lcuspy
	\end{cases} \per
\end{align}
In Case Ia gravitational radiation controls the loop decay ($dL/dt \approx -\Gamma_g G M^2$), in Case II particle radiation controls the loop decay ($dL/dt \approx - \Gcusp / (mL)^{1/2}$), and in Case Ib gravitational radiation controls the decay while the loop is large ($L \gg \Lcuspy$) and particle radiation controls the decay once the loop becomes small ($L \ll \Lcuspy$).  
We determine the loop lifetime, $\tdecay$, by solving $L(\tdecay + t_{i}) = 0$ to obtain
\begin{align}\label{eq:x_decay}
	\tdecay 
	\approx \begin{cases}
	\frac{L_i}{\Lcuspy} \tcuspy
	& \Lcuspy \ll L_{i} \\
	\frac{2}{3} \left( \frac{L_i}{\Lcuspy} \right)^{3/2} \tcuspy
	& L_{i} \ll \Lcuspy
	\end{cases} 
\end{align}
where the upper expression is for both Cases Ia and Ib.  

We study loop decay here so that we may calculate the distribution of loop lengths for the entire string network in \sref{sub:LoopDensity}.  
In the literature one often assumes an instantaneous decay approximation, that is $L \approx L_i$ for $t < \tdecay$ and $L = 0$ afterward.  
In this approximation, the network is devoid of loops smaller than a minimum length determined by solving $t = \tdecay$ for $L_i$.  
Observables associated with the string network are calculated by integrating over loop length.  
The approximation proves to be a very good one provided that the integral is not sensitive to small $L$ as in the case of radiation from cusps where the power goes as $1 / \sqrt{L}$.
Since the particle radiation power output from kinky loops grows like $1/L$ with decreasing $L$, it is not {\it a priori} clear that the instantaneous decay approximation is sufficient for us.

\begin{figure}[t]
\hspace{0pt}
\vspace{-0in}
\begin{center}
\includegraphics[width=0.49\textwidth]{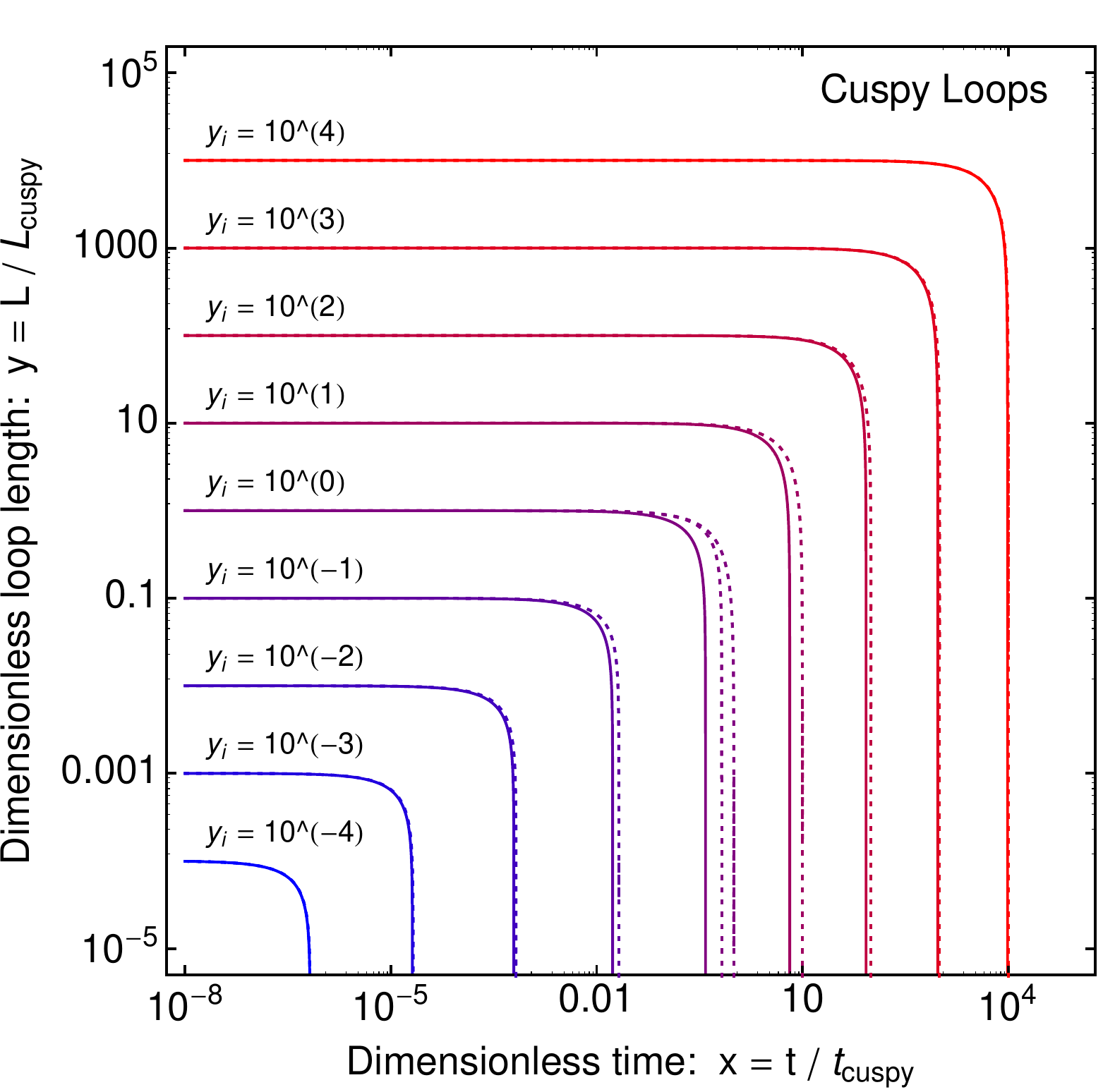} \hfill
\includegraphics[width=0.49\textwidth]{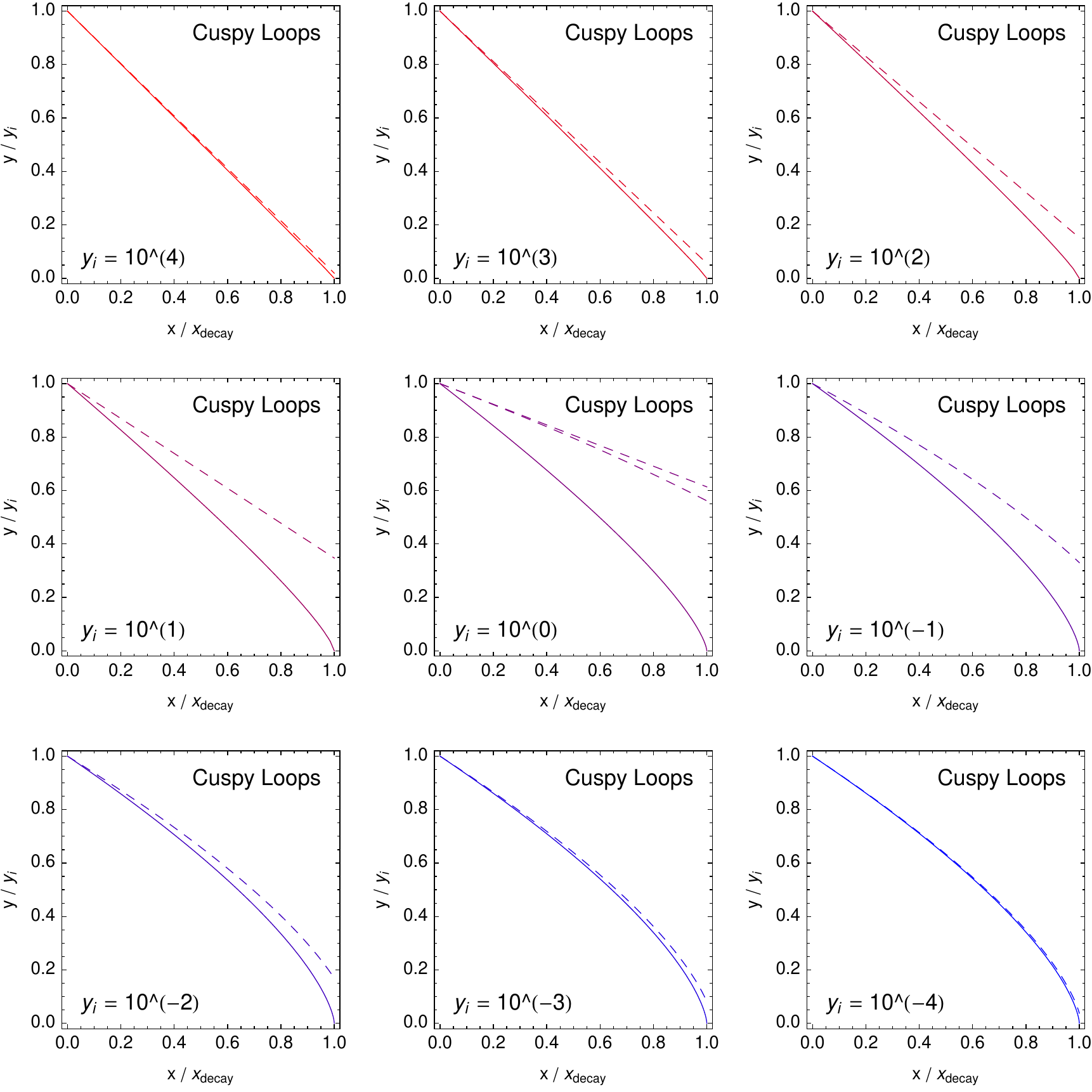} 
\caption{
\label{fig:loopevolve}
{\it Left:}
The solution of \eref{eq:yeqn_cuspy} determined numerically (solid) and the approximate solution (dotted) in \eref{eq:Loft_cuspy}.  There are two dashed lines for $y_i = 1$ where we show both terms in \eref{eq:Loft_cuspy}.  
In the instantaneous decay approximation, these curves would be replaced with a step function.
{\it Right:}  
Same analysis presented on a linear scale where $y/y_i$ is the fraction of the initial loop length and $x / x_{\rm decay}$ is the fraction of the total loop lifetime ($x_{\rm decay} = t_{\rm decay} / \tcuspy$).  
}
\end{center}
\end{figure}

We will generalize the instantaneous decay approximation in the following way.  
We saw in \eref{eq:ysol_cuspy_1} that the loop decay is controlled by particle radiation if $L_i \ll L_{\rm cuspy}$.  
We will suppose that gravitational radiation is responsible for loop decay if $L_i \gg \Lcuspy$, {\it even if} $L < \Lcuspy$.  
Then we can take Cases Ia and II from \eref{eq:ysol_cuspy_1} and insert $\Lcuspy$ and $\tcuspy$ from \erefs{eq:L_cuspy}{eq:t_cuspy} to write the loop length as 
\begin{align}\label{eq:Loft_cuspy}
	L(t) & \approx 
	\Bigl[ L_i - \Gamma_g G M^2 (t - t_i) \Bigr]
	\Theta(L_i - \Lcuspy) 
	+ \Bigl[ L_i^{3/2} - \frac{3}{2} \Gcusp \frac{t-t_i}{\sqrt{m}} \Bigr]^{2/3}
	\Theta(\Lcuspy - L_i) \per
\end{align}
where $\Theta$ is the Heaviside step function.  
Here we are really making two approximations:  we neglect the solutions that fall into the category of Case Ib ($L < \Lcuspy < L_i$), and we assume that the transition between the asymptotic behaviors is abrupt.  
Both approximations are justified because we are ultimately interested in integrated quantities.  
Upon integrating over $L$ and $t$, the contributions from $L \ll \Lcuspy \ll L_i$ and $L_i \approx \Lcuspy$ are not significant.  
We further clarify these points in \sref{sub:LoopDensity}.  
In \fref{fig:loopevolve} we show the exact solution of \eref{eq:yeqn_cuspy} (determined numerically) and the approximate solutions in \eref{eq:Loft_cuspy}.  
The approximation works very well except when $L_i \approx \Lcuspy$ ($y_i \approx 1$).  

The loop lifetime is obtained from \eref{eq:x_decay}:  
\begin{align}\label{eq:tau_est}
	\tau \approx 
	\frac{L_i}{\Gamma_g G M^2} 
	\Theta(L_i - \Lcuspy)
	+ \frac{2}{3} \frac{L_i^{3/2} \sqrt{m}}{\Gcusp}
	\Theta(\Lcuspy - L_i) \per
\end{align}
The loop oscillates with a frequency $f \sim 1 / L$, and therefore it experiences $\mathcal{N} \approx \tdecay / L_i$ oscillations before it decays.  
For a wide range of parameters, ${\rm TeV} < M < 10^{18} \GeV$ and $L_i \gtrsim 1/M$, we have $\mathcal{N} \gg 1$, and the loop oscillates many times.  
This estimate is a self-consistency check:  the radiation powers presented in \sref{sec:SM_Particle_Prod} are derived assuming that the loop is oscillating periodically, and that any transient behavior has died away.  
\eref{eq:tau_est} can also be written as 
\begin{align}
	\tdecay \approx 
	\tcuspy \frac{L_i}{\Lcuspy}
	\Theta(L_i - \Lcuspy)
	+ \tcuspy \frac{2}{3}  \frac{L_i^{3/2}}{\Lcuspy^{3/2}}
	\Theta(\Lcuspy - L_i) \com
\end{align}
which illustrates that $\tcuspy$ is the lifetime of a loop of initial length $L_{i} \approx \Lcuspy$; larger loops are longer lived $\tdecay \gg \tcuspy$ and smaller loops decay more quickly $\tdecay \ll \tcuspy$.  

\qquad \\
{\bf (ii) Kinky Loops} \\
We can perform a similar analysis for the case of kinky loops.  
The loop decay equation takes the form\footnote{To our knowledge, \rref{Lunardini:2012ct} contains the only study of kinky loops decaying both gravitationally and by particle emission.  However, they take the particle radiation power to scale as $P \sim L^{0}$ whereas we have argued $P \sim L^{-1}$ in \rref{Long:2014mxa}, and consequently our loop decay equation is of a different form.  }
\begin{align}\label{eq:dLdt_kinky}
	M^2 \frac{dL}{dt} = - \Gamma_{g} G M^4 - M^2 \frac{\Gkk}{mL} 
	\qquad , \qquad
	L(t_i) = L_i
\end{align}
where we sum the powers due to gravitational radiation and particle emission from kink collisions, \erefs{eq:Pgrav}{eq:P_kk}, on the right hand side.  
The characteristic length and time scale are 
\begin{align}
	\Lkinky & \equiv \frac{\Gkk}{\Gamma_g G} \frac{1}{M^2 m} \label{eq:L_kinky} \\
	\tkinky & \equiv \frac{ \Gkk }{(\Gamma_g G)^2 M^4 m} \label{eq:t_kinky} \per
\end{align}
Since $G M^2 \ll 1$, we have $\Lkinky \ll \Lcuspy$ and $\tkinky \ll \tcuspy$.  
The approximate solution is 
\begin{align}\label{eq:Loft_kinky}
	L(t) & \approx 
	\bigl[ L_i - \Gamma_g G M^2 (t - t_i) \bigr] \Theta( L_i - \Lkinky )
	+ \bigl[ L_i^{2} - 2 \frac{\Gkk}{m} (t - t_i) \bigr]^{1/2} \Theta( \Lkinky - L_i ) \per
\end{align}
The loop lifetime is given by 
\begin{align}
	\tau \approx 
	\frac{L_i}{\Gamma_g G M^2} 
	\Theta(L_i - \Lkinky)
	+ \frac{1}{2} \frac{L_i^{2} m}{\Gkk}
	\Theta(\Lkinky - L_i) \per
\end{align}

\subsection{Loop Density Function}\label{sub:LoopDensity}

At any given time, the string network will be populated by loops of different sizes as new loops form and existing loops decay.  
Let
\begin{align}
	dn_{L} = \nu(t,L) dL
\end{align}
be the number density of loops (per unit physical volume) with length between $L$ and $L + dL$ at time $t$.  
We calculate $\nu(t,L)$ using empirical input from string network simulations.  
One might expect that the simulations would tell us $\nu(t,L)$ directly, but this is not the case.  
Nambu-Goto string simulations neglect the radiative processes responsible for loop decay, and this affects $\nu(t,L)$ since a small loop at time $t$ was formed as a larger loop at an earlier time.  
Instead, the simulations provide us with the empirical loop formation rate as a function of length.  

To calculate $\nu(t,L)$ we must also know the loop length flow $L_i(t_i; L, t)$, {\it i.e.} the length of a loop at time $t_i$ that later has a smaller length $L$ at time $t$.  
As we saw in \sref{sub:LoopDecay}, it is not always possible to obtain an exact expression for the loop length flow.  
Specifically, one cannot solve \eref{eq:yeqn_cuspy} for $y_i = L_i / \Lcuspy$ in closed form.  
The difficulty is that radiation is being emitted in two different channels (gravitational and particle emission) at rates that depend on the loop length in different ways.  
However, in the approximations leading to \erefs{eq:Loft_cuspy}{eq:Loft_kinky} we saw that we can consider separately the cases in which either gravitational or particle radiation is dominant.  
In these two regimes, the loop length flow is found by solving the loop decay equation, \eref{eq:loop_decay_eqn}, with a power of the form $P = M^2 \pp / (mL)^{\idx}$.  
If gravitational radiation dominates then $\pp = \Gamma_g G M^2$ and $\idx = 0$, and if particle radiation dominates then $\pp = \Gcusp$ and $\idx = 1/2$ for cuspy loops or $\pp = \Gkk$ and $\idx = 1$ for kinky loops.  

In \aref{app:loop_dist} we calculate $\nu(t,L)$ as described above.  
The results are given by \eref{eq:n_physical} and reproduced here:
\begin{align}\label{eq:loop_dist}
	\nu(t,L) = \left\{
	\begin{array}{lcl}
	\oef \, 
	\frac{1}{t^{3/2}} \, 
	\frac{L^{\idx} }{L_0(t,L)^{5/2+\idx}} \,
	\Theta(1- \frac{L}{0.1 t}) \, 
	& \quad &
	t < t_{eq} 
	\\
	\oef \, 
	\frac{t_{eq}^{1/2}}{t^2} \, 
	\frac{L^{\idx}}{ L_0(t,L)^{5/2 + \idx}} \,
	\Theta(1 - \frac{L_{eq}(t,L)}{0.1 t_{eq}}) 
	& \quad & 
	t > t_{eq} 
	\\
	\quad + \, 
	\tss
	\Bigl[ 1 - \bigl( \frac{L}{0.18 t} \bigr)^{0.31} \Bigr] \, 
	\frac{1}{t^2} \, 
	\frac{L^{\idx} }{L_0(t,L)^{2+\idx} } 
	& \quad & 
	\end{array} 
	\right.
\end{align}
where $\Theta(x)$ is the Heaviside step function.  
The loop formation rate differs in the radiation $(t < t_{eq})$ and matter eras $(t > t_{eq})$ leading to the two cases.  
The functions 
\begin{align}
	\label{eq:Leq}
	L_{eq}(t,L) & = \bigl[ L^{\idx+1} + (1+\idx) \frac{\pp}{m^{\idx}} (t-t_{eq}) \bigr]^{1/(\idx+1)} \com
	\\
	\label{eq:L0}
	L_0(t,L) & = \bigl[ L^{\idx+1} + (1 + \idx) \frac{\pp}{m^{\idx}} t \bigr]^{1 / (\idx +1)} 
\end{align}
are the lengths of a loop at RM equality and at $t=0$, respectively, that later has length $L$ at time $t$.  

In deriving \eref{eq:loop_dist} we assumed that loop creation was continuous since the formation of the string network.  
As we discussed in \sref{sub:Friction}, however, loops that formed during the friction-dominated epoch ($t < t_{\ast}$) were over-damped and decayed away quickly.  
Then one should only integrate back to the end of the friction era in calculating $\nu(t,L)$, and doing so leaves a deficit of loops smaller than $L \sim 0.1 t_{\ast}$.  
In writing \eref{eq:loop_dist} we implicitly assume that friction domination ends sufficiently early such that $t_{\ast}$ is much smaller than the loop length scales of interest.  
If friction domination lasts all the way until the epoch of $e^+e^-$ annihilations then $t_{\ast} \sim 10^{2} \sec$, and the abundance of loops smaller than $L \sim 10^{6} \km$ is suppressed.  

During the matter era ($t > t_{eq}$), the cosmic string network consists of two populations: the relic loops that survived from the radiation era and the loops that were newly formed in the matter era.  
These populations correspond respectively to the two terms in $\nu(t > t_{eq},L)$ above, and they have the ratio \begin{align}\label{eq:lr_ov_lm}
	\frac{\text{R-era relic loops}}{\text{new M-era loops}}
	\approx \frac{t_{eq}^{1/2}}{L_0(t,L)^{1/2}} 
	\gg 1 
\end{align}
for $L_{eq}(t,L) < 0.1 t_{eq} \approx 10^{17} \km$.  
For the calculation of observables in \sref{sec:Observables} we will primarily be interested in much smaller loops, since the particle radiation power grows with decreasing loop size (see \erefs{eq:P_cusp}{eq:P_kk}).  
Therefore we are well-justified in keeping only the R-era relic loops (first term of $\nu(t > t_{eq},L)$).  
However, to keep the expressions general, we retain both terms in this section.  

Note the factor of $(L / L_0)^{\idx} \leq 1$ in \eref{eq:loop_dist}, and recall that the index $\idx$ controls the loop's rate of energy loss via $P \propto L^{-\idx}$.  
For gravitational radiation ($\idx = 0$) the factor reduces to unity, but for a loop that decays predominately by particle emission ($\idx \neq 0$) the factor suppresses the abundance of small loops ($L \ll L_0$).  
Large loops ($L \lesssim L_0$) are unaffected as they have not yet had time to decay.  
This factor has not been included in previous calculations of the loop distribution, even in the regime where particle emission is dominant ($\idx \neq 0$).  
As discussed in \aref{app:loop_dist}, the factor arises from the radiation power's dependence on the loop length:  
loops with length between $L$ and $L + dL$ at time $t$ arose from a population of loops in a narrower 
band of lengths at an earlier time:  $dL_{i} = (\partial L_{i} / \partial L) dL = (L / L_{i})^{\idx} dL$ with 
$(L / L_{i})^{\idx} \leq 1$.  

Using \eref{eq:loop_dist} we can find the loop distribution that results from different radiation processes:  
gravitational radiation ($\idx = 0$, $\pp = \Gamma_g G M^2$),
\begin{align}\label{eq:loop_dist_grav}
	\nu_g(t,L) = \left\{
	\begin{array}{lcl}
	\oef \, 
	\frac{1}{t^{3/2}} \, 
	\frac{1}{ [L + \Gamma_g G M^2 t ]^{5/2}} \,
	\Theta( 1 - \frac{L}{0.1 t} )
	& \qquad & 
	t < t_{eq} \\
	\oef \, 
	\frac{t_{eq}^{1/2}}{t^2} \, 
	\frac{1}{ [ L + \Gamma_g G M^2 t ]^{5/2}} \,
	\Theta(1 - \frac{ L + \Gamma_g G M^2 (t - t_{eq}) }{0.1 t_{eq}}) 
	& \qquad & 
	t > t_{eq} \\
	\quad + \, 
	\tss
	\Bigl[ 1 - \bigl( \frac{L}{0.18 t} \bigr)^{0.31} \Bigr] \, 
	\frac{1}{t^2} \, 
	\frac{1}{[ L + \Gamma_g G M^2 t ]^{2} } 
	\end{array} \right. \com
\end{align}
Particle radiation from cusps ($\idx = 1/2$, $\pp = \Gcusp$), 
\begin{align}\label{eq:loop_dist_cusp}
	\nu_c(t,L) = \left\{
	\begin{array}{lcl}
	\oef \, 
	\frac{1}{t^{3/2}} \, 
	\frac{L^{1/2} }{\bigl[ L^{3/2} + \frac{3}{2} \Gcusp \frac{t}{m^{1/2}} \bigr]^{2}} \,
	\Theta( 1 - \frac{L}{0.1 t} )
	& \qquad & 
	t < t_{eq} \\
	\oef \, 
	\frac{t_{eq}^{1/2}}{t^2} \, 
	\frac{L^{1/2}}{ \bigl[ L^{3/2} + \frac{3}{2} \Gcusp \frac{t}{m^{1/2}} \bigr]^{2}} \,
	\Theta(1 - \frac{ L^{3/2} + \frac{3}{2} \Gcusp \frac{t-t_{eq}}{m^{1/2}} }{0.03 \, t_{eq}^{3/2} }) 
	& \qquad & 
	t > t_{eq} \\
	\quad + \, 
	\tss
	\Bigl[ 1 - \bigl( \frac{L}{0.18 t} \bigr)^{0.31} \Bigr] \, 
	\frac{1}{t^2} \, 
	\frac{L^{1/2} }{\bigl[ L^{3/2} + \frac{3}{2} \Gcusp \frac{t}{m^{1/2}} \bigr]^{5/3} } 
	\end{array} \right. \com
\end{align}
and particle radiation from kink collisions ($\idx = 1$, $\pp = \Gkk$),
\begin{align}\label{eq:loop_dist_kk}
	\nu_{kk}(t,L) = \left\{
	\begin{array}{lcl}
	\oef \, 
	\frac{1}{t^{3/2}} \, 
	\frac{L }{\bigl[ L^{2} + 2 \Gkk \frac{t}{m} \bigr]^{7/4}}
	\Theta( 1 - \frac{L}{0.1 t} )
	& \qquad & 
	t < t_{eq} \\
	\oef \, 
	\frac{t_{eq}^{1/2}}{t^2} \, 
	\frac{L }{\bigl[ L^{2} + 2 \Gkk \frac{t}{m} \bigr]^{7/4}}
	\Theta(1 - \frac{ L^{2} + 2 \Gkk \frac{t-t_{eq}}{m} }{0.01 \, t_{eq}^2 }) 
	& \qquad & 
	t > t_{eq} \\
	\quad + \, 
	\tss
	\Bigl[ 1 - \bigl( \frac{L}{0.18 t} \bigr)^{0.31} \Bigr] \, 
	\frac{1}{t^2} \, 
	\frac{L }{\bigl[ L^{2} + 2 \Gkk \frac{t}{m} \bigr]^{3/2}}
	\end{array} \right. \per
\end{align}

As discussed in \sref{sub:LoopDecay}, the rate of energy loss from a loop of length $L$ is not generally of the form $P = \pp M^2 / (mL)^{\idx}$, as we assumed in the derivation of \eref{eq:loop_dist}.  
Loops that are large decay predominantly via gravity wave emission ($\idx = 0$), and those that are small decay primarily through particle emission ($\idx \neq 0$).  
However, we saw that the solution $L(t)$ can be approximated as in \erefs{eq:Loft_cuspy}{eq:Loft_kinky}, that is, particle emission can be neglected if the loop was large at formation ($L_0 > \Lcuspy$ or $\Lkinky$) and gravitational emission can be neglected if the loop was small at formation ($L_0 < \Lcuspy$ or $\Lkinky$).  
The total loop length distribution is obtained by summing these two contributions.  
For cuspy loops we have 
\begin{align}\label{eq:nu_cuspy}
	\nu_{\rm cuspy}(t,L) = f_c \Bigl[ &
	\nu_g(t,L) \ \Theta \bigl( L + \Gamma_g G M^2 t - \Lcuspy \bigr) 
	+ \nu_c(t,L) \ \Theta \bigl( \Lcuspy - L - \Gamma_g G M^2 t \bigr)
	\Bigr]
\end{align}
and for kinky loops we obtain 
\begin{align}\label{eq:nu_kinky}
	\nu_{\rm kinky}(t,L) = (1 - f_c) \Bigl[ &
	\nu_g(t,L) \, \Theta \bigl( L + \Gamma_g G M^2 t - \Lkinky \bigr) 
	+ \nu_{kk}(t,L) \, \Theta \bigl( \Lkinky - L - \Gamma_g G M^2 t \bigr)
	\Bigr] \per
\end{align}
Here $f_c \leq 1$ is the fraction of loops that have cusps.  
When backreaction is neglected, numerical study suggest that $f_c \approx 0.4$ \cite{Copi:2010jw};
the effects of backreaction on $f_c$ are not known.

\begin{figure}[t]
\hspace{0pt}
\vspace{-0in}
\begin{center}
\includegraphics[width=0.49\textwidth]{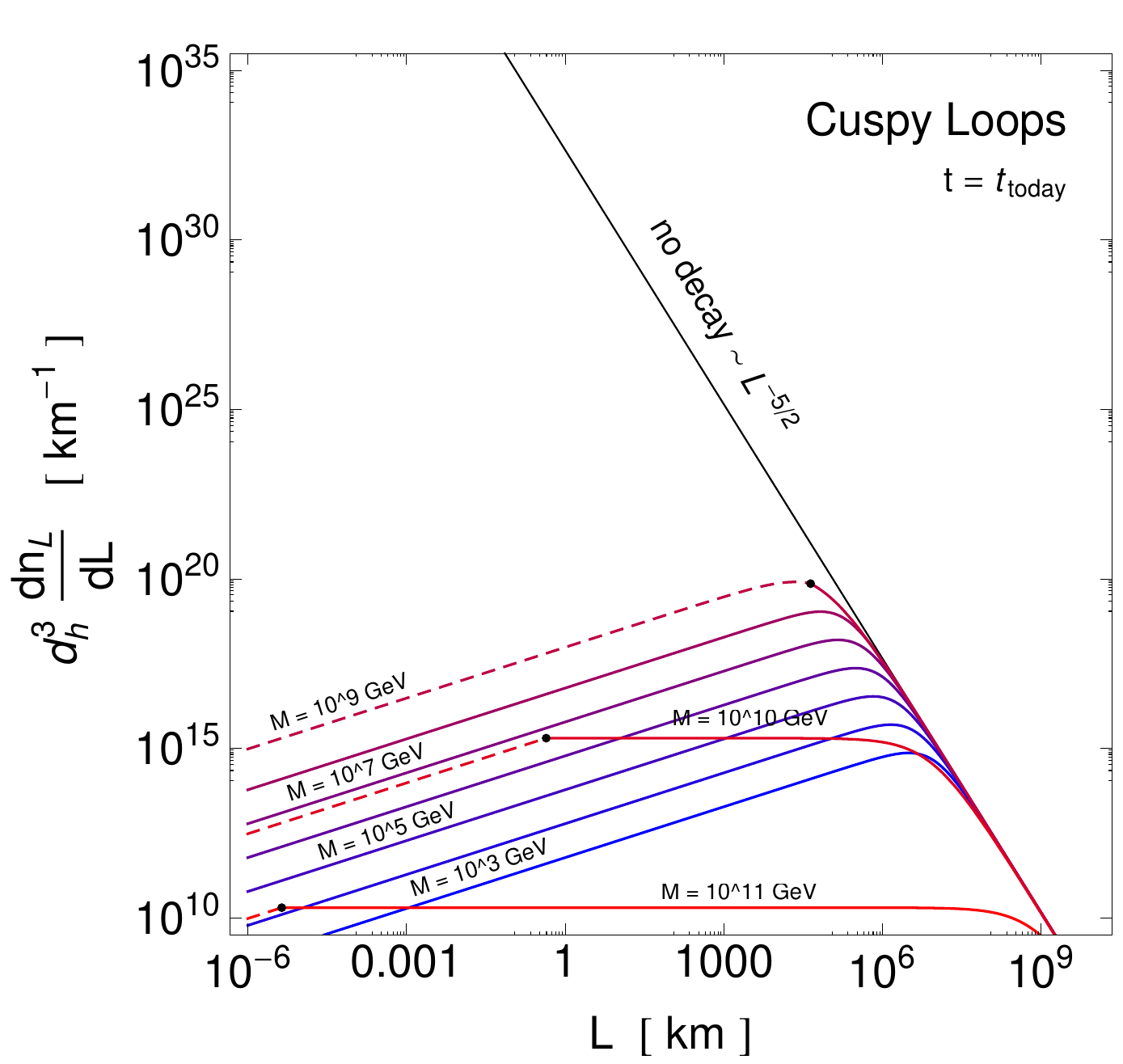} 
\includegraphics[width=0.49\textwidth]{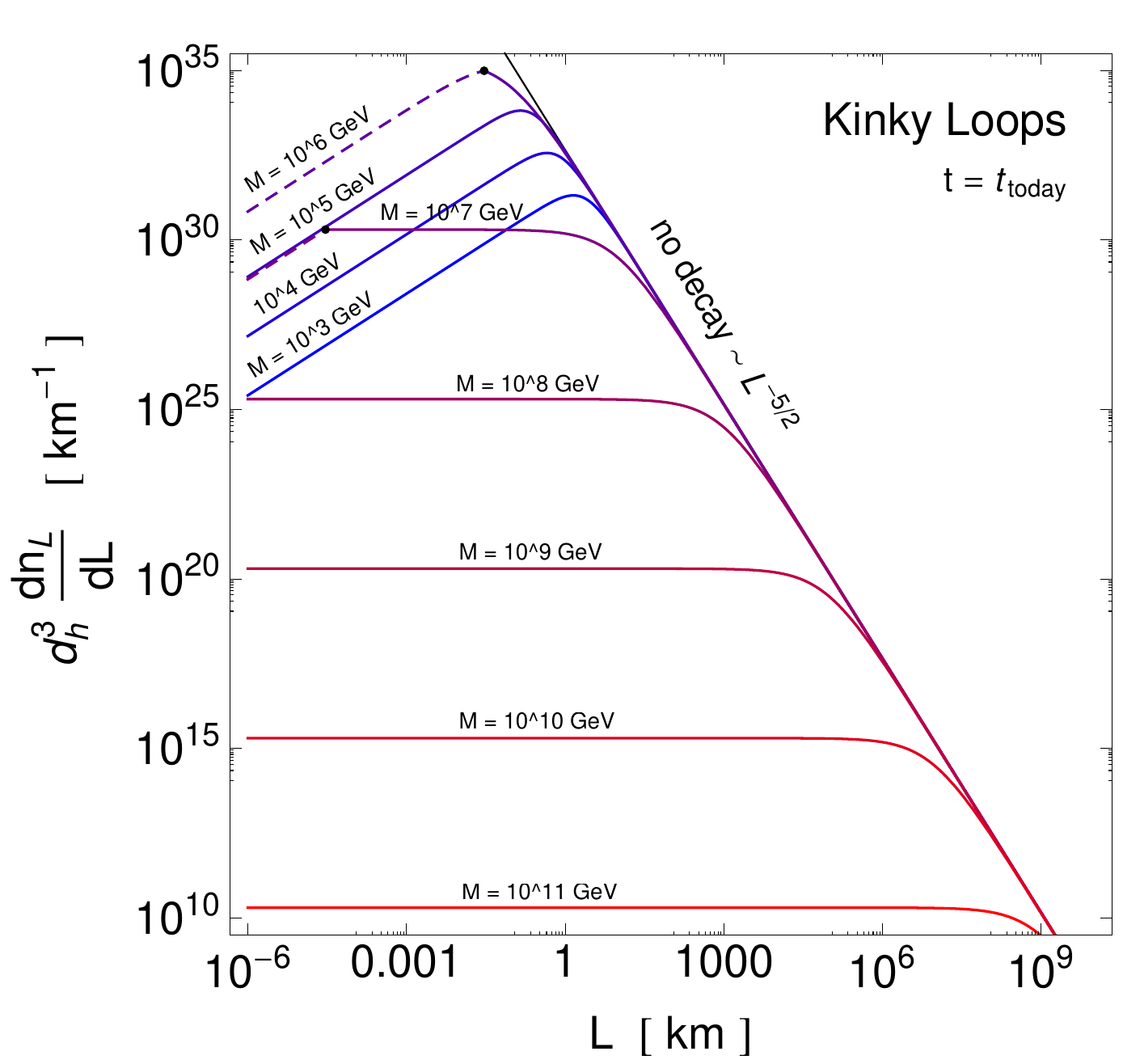} 
\caption{
\label{fig:looplength}
The number of loops per unit length that are within the observable universe today ($d_h = 46.5 \times 10^{9} \, {\rm ly}$; \ $t_{\rm today} = 13.7 \times 10^{9} \, {\rm yr}$) for models with different values of the string tension ($M = \sqrt{\mu}$).  
The left panel is for cuspy loops, and the right panel for kinky loops.  
To make these figures, we have expressed the effective couplings $\Gcusp$ and $\Gkk$ in terms of $\gHHstr$ using \eref{eq:Gamma_HH}, and we have fixed $\gHHstr = 1$.
}
\end{center}
\end{figure}

The loop length distribution in the present era ($t_{\rm today} \approx 13.7 \, {\rm Gyr}$) is plotted in \fref{fig:looplength} using \erefs{eq:nu_cuspy}{eq:nu_kinky}.  
The dimensionless coefficients are written in terms of $\gHHstr$ using \eref{eq:Gamma_HH} where we take the upper range of the coefficient; we take $\gHHstr = 1$, and we show various models with different string tensions $\mu = M^2$.  
The line color and labels are intended to highlight the non-monotonic behavior of the loop abundance with varying $M$, which was first observed in \rref{Bhattacharjee:1989vu}.  
For large $M$ the gravitational radiation is so efficient that all of the loops today were large at formation ($L + \Gamma_g G M^2 t_{\rm today} > \Lcuspy$ or $\Lkinky$); the loops that were small at formation have decayed away entirely.  
In the notation of \sref{sub:LoopDecay} this translates into $t_{\rm today} > \tcuspy$ or $\tkinky$.  
Using \erefs{eq:t_cuspy}{eq:t_kinky} this implies $M > M_{\rm cuspy}(t_{\rm today})$ or $M_{\rm kinky}(t_{\rm today})$ where $M_{\rm cuspy}(t)$ and $M_{\rm kinky}(t)$ are the solutions to the equations
\begin{align}
	M_{\rm cuspy} = \Bigl[ \frac{[\Gcusp]^2}{\Gamma_g^3 G^3 m t} \Bigr]^{1/6} 
	\qquad {\rm and} \qquad
	M_{\rm kinky} = \Bigl[ \frac{\Gkk}{\Gamma_g^2 G^2 m t} \Bigr]^{1/4} \label{eq:M_cuspy_kinky} 
	\per
\end{align}
Note that $\Gcusp$ and $\Gkk$ may also depend on $M$ as in \erefs{eq:Gamma_HH}{eq:Gamma_H}.  
For the parameters used in \fref{fig:looplength} we have $M_{\rm cuspy}(t_{\rm today}) \approx 9 \times 10^{8} \GeV$ and $M_{\rm kinky}(t_{\rm today}) \approx 5 \times 10^{5} \GeV$.  
For $M \approx M_{\rm cuspy}(t_{\rm today})$ or $M_{\rm kinky}(t_{\rm today})$ the total loop abundance $\int dn_L$ is maximal.

In the regime of large $M > M_{\rm cuspy}(t_{\rm today})$ or $M_{\rm kinky}(t_{\rm today})$, we have $\nu_{\rm cuspy} / f_c \approx \nu_{\rm kinky} / (1 - f_c) \approx \nu_g \propto L_0^{-5/2} = (L + \Gamma_g G M^2 t)^{-5/2}$ over the entire range of loop lengths shown.  
For large loops, the distribution falls like $L^{-5/2}$, since these loops have not yet had sufficient time to decay appreciably and $L_0 \approx L$.  
For small loops, the length distribution function is suppressed due to decay and becomes independent of $L$.  
Below $L = \Lcuspy$ or $\Lkinky$, denoted by a black dot, the assumptions that go into \erefs{eq:nu_cuspy}{eq:nu_kinky} break down; this is discussed further below.  
The most abundant loops are those that are just beginning to decay today.  
These loops have a length $L_{\rm peak} \approx \Gamma_g G M^2 t_{\rm today}$, and on \fref{fig:looplength} this corresponds to the point of transition between the $L^{-5/2}$ and $L^{0}$ scaling.  

\begin{figure}[t]
\hspace{0pt}
\vspace{-0in}
\begin{center}
\includegraphics[width=0.49\textwidth]{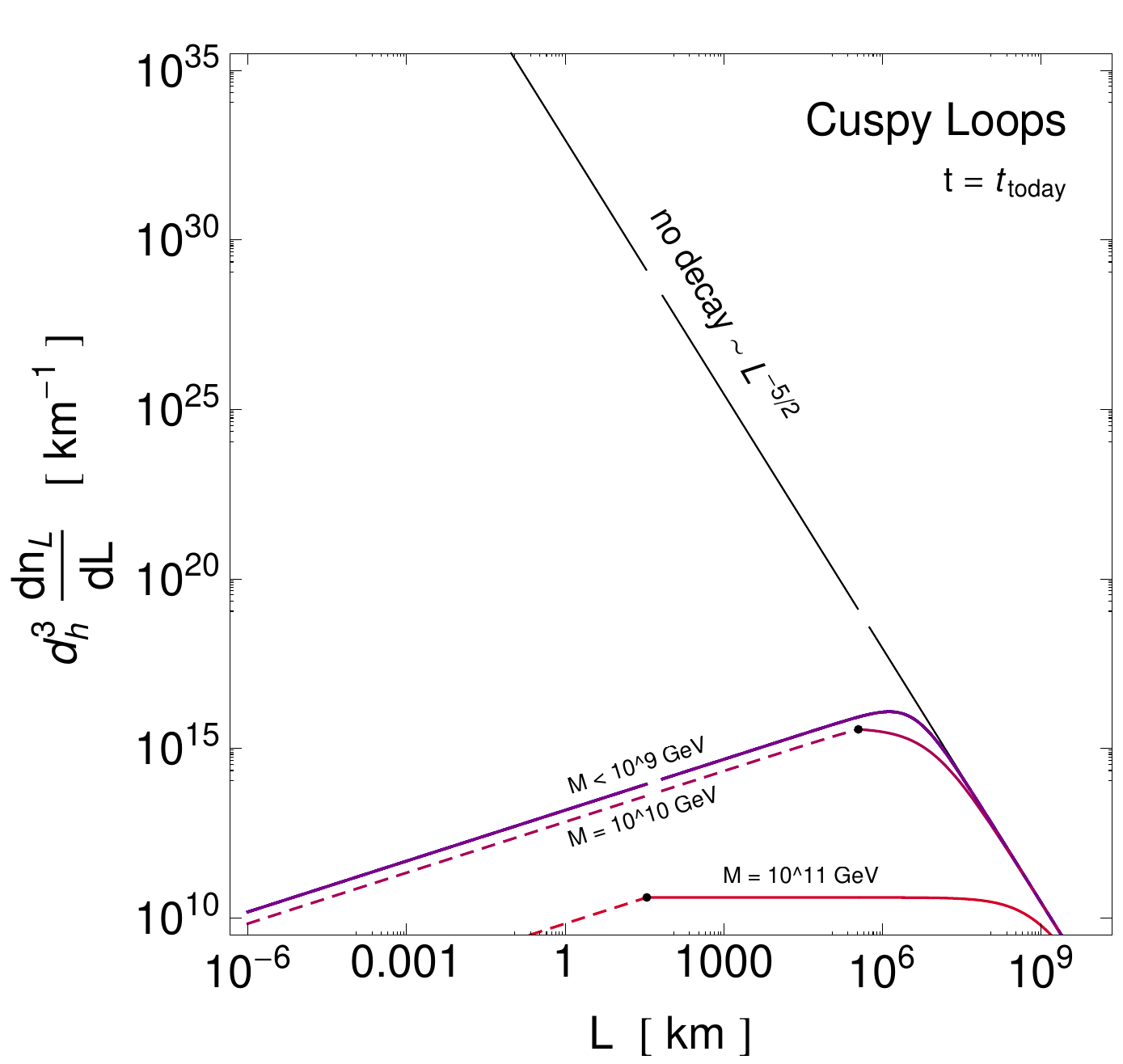} 
\includegraphics[width=0.49\textwidth]{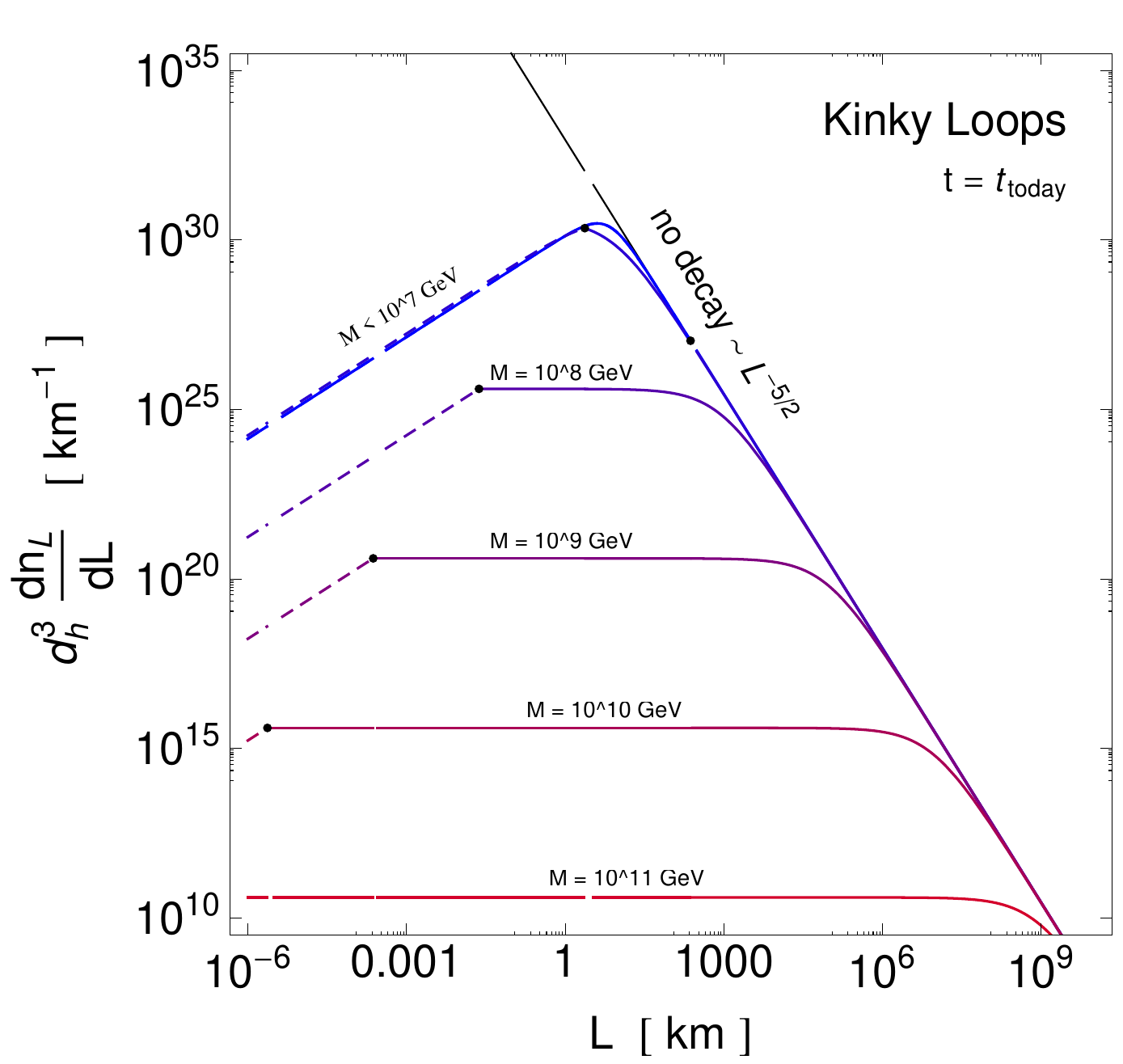} 
\caption{
\label{fig:looplength_2}
The loop length distribution for the case where $\Gcusp$ and $\Gkk$ are given by \eref{eq:Gamma_H}.  
We have fixed $\gHstr = 1$ and $\Hcond = 0.1 M$.
}
\end{center}
\end{figure}

In the regime of small $M$ we have $t_{\rm today} < \tcuspy$ or $\tkinky$.  
Here the gravitational radiation is so inefficient that loops which were large at formation ($L_0 = L + \Gamma_g G M^2 t_{\rm today} > \Lcuspy$) are still large ($L \approx L_0$) today; all of the loops that are small today ($L < \Lcuspy$) were also small at formation ($L_0 < \Lcuspy$).  
At large $L$ where $L \approx L_0$ we have $\nu \sim L^{-5/2}$, just as in the large $M$ case discussed above. 
At small $L$, the length distribution is suppressed by loop decay.  
The scaling would go as $\nu \sim L^{0}$, but the additional Jacobian factor, which was discussed above  \eref{eq:loop_dist_grav}, {\it further suppresses} the loop abundance and leads to the scalings $\nu_{\rm cuspy} \sim L^{1/2}$ and $\nu_{\rm kinky} \sim L^{1}$.  
The most abundant loops have a length $L_{\rm peak} \approx \bigl[ \frac{3}{2} \Gcusp \frac{t_{\rm today}}{m^{1/2}} \bigr]^{2/3}$ (cuspy) or $L_{\rm peak} \approx \bigl[ 2 \Gkk \frac{t_{\rm today}}{m} \bigr]^{1/2}$ (kinky) such that they are just decaying today.  

In the region where the curves are dashed, the assumptions that go into the derivation of \erefs{eq:nu_cuspy}{eq:nu_kinky} have broken down.  
Specifically, the derivation assumes that the loop decay is dominated by gravitational radiation at all length scales provided that the {\it initial} length is large, $L_0 = L + \Gamma_g G M^2 t > \Lcuspy$ or $\Lkinky$.  
(In the notation of \eref{eq:length_cases}, we subsumed Case IIb into Case Ia.)  
More realistically, particle radiation dominates once the large loops become small.  
In this regime, the Jacobian factor $\partial L_{i} / \partial L$ will further suppress the loop abundance, as discussed above.  
The appropriate factor is obtained from the middle case of \eref{eq:ysol_cuspy_1} to be $\partial L_i / \partial L = (L / \Lcuspy)^{1/2}$ for cuspy loops and $(L / \Lkinky)^{1}$ for kinky loops.  
The dashed curves in \fref{fig:looplength} have been scaled by these factors.  

Whereas \fref{fig:looplength} shows the loop length distribution when $\Gcusp$ and $\Gkk$ are given by \eref{eq:Gamma_HH}, we also show in \fref{fig:looplength_2} the length distribution for $\Gcusp$ and $\Gkk$ given by \eref{eq:Gamma_H}.  
By imposing $\Hcond \propto M$, the couplings $\Gcusp \approx \Gkk \approx (\gHstr)^2 (\Hcond / M)^2$ become independent of $M$.  
In this case the loop length distributions also become independent of $M$ when particle emission is dominant, and $\nu = \nu_c$ or $\nu_{kk}$ as in \erefs{eq:loop_dist_cusp}{eq:loop_dist_kk}.

\subsection{Higgs Injection Function}\label{sub:Higgs_Injection}

Particle emission from string loops, which is dominantly in the form of Higgs bosons, plays a central role in the various astrophysical and cosmological probes of the string network.
The quantity of interest is $\Pcal_{H}(t) dt$, that is, the energy that is ejected from the string network in the form of Higgs bosons between time $t$ and $t+dt$ and per unit physical volume.  
By summing the populations of cuspy and kinky loops and integrating over loop length we have the Higgs injection function \cite{Vachaspati:2009kq} 
\begin{align}\label{eq:PH_def}
	\Pcal_{H}(t) 
	& = \int_{0}^{\infty} \, \Bigl( \nu_{\rm cuspy}(t,L) \, \Pcusp(L) + \nu_{\rm kinky}(t,L) \, \Pkk(L) \Bigr) dL 
\end{align}
where the Higgs emission powers are given by \erefs{eq:P_cusp}{eq:P_kk} and the loop length distributions are given by \erefs{eq:nu_cuspy}{eq:nu_kinky}.  

One calculates $\Pcal_{H}(t)$ by performing the integral over loop length $L$.  
Since the loop length distributions are given by piecewise functions, \erefs{eq:nu_cuspy}{eq:nu_kinky}, we break up the integration domain into small loops and large loops.  
For the large loops we use $\nu_{\rm cuspy} \approx \nu_{\rm kinky} \approx \nu_g$ and for the small loops it is $\nu_{\rm cuspy} \approx \nu_{c}$ or $\nu_{\rm kinky} \approx \nu_{kk}$.  
The demarcation point between large and small is determined by $L + \Gamma_g G M^2 t \gtrless \Lcuspy$ or $\Lkinky$, which is a function of time $t$.  
The time variable also must be integrated to obtain the various observables of interest.  
Although this calculation is straightforward, it is tedious to keep track of the limits of integration and ultimately unnecessarily.  

In an alternative approach, one performs two separate calculations:  first assuming that gravitational radiation has the dominant influence on the loop decay rate {\it at all length scales} ($\nu_{\rm cuspy} \approx \nu_{\rm kinky} \approx \nu_g$), then assuming that particle radiation dominates at all length scales ($\nu_{\rm cuspy} \approx \nu_{c}$ or $\nu_{\rm kinky} \approx \nu_{kk}$), and finally $\Pcal_{H}$ is estimated as the smaller of two outcomes \cite{Dufaux:2012np}.  
This approximation is written as 
\begin{align}\label{eq:PH_piecewise}
	\Pcal_H(t) = \ & 
	f_{c} \times {\rm Min} \left\{
	\begin{array}{l}
	\int_{0}^{\infty} dL \, \nu_{g}(t,L) \, \Pcusp(L)
	\\
	\int_{0}^{\infty} dL \, \nu_{c}(t,L) \, \Pcusp(L)
	\end{array} \right.
	\nn
	& + (1 - f_{c}) \times {\rm Min} \left\{
	\begin{array}{l}
	\int_{0}^{\infty} dL \, \nu_{g}(t,L) \, \Pkk(L)
	\\
	\int_{0}^{\infty} dL \, \nu_{kk}(t,L) \, \Pkk(L)
	\end{array} \right. 
\end{align}
where ${\rm Min} \bigl\{ a , b  = a$ if $a < b$ and $b$ if $ b \leq a$.  
Relaxing the limits of integration necessarily overestimates the integral (the integrand is always positive), and this is why we must take the smaller of the two expressions.  
We can evaluate \eref{eq:PH_piecewise} separately in the radiation and matter eras to find:
\begin{align}\label{eq:PH_approx}
	\Pcal_H & (t < t_{eq}) \approx 
	\frac{t^{1/2}} {t_{eq}^{1/2}} \Pcal_H(t > t_{eq}) 
	\nn & 
	\approx f_{c} \times {\rm Min} \left\{
	\begin{array}{l}
	\tff \frac{\Gcusp}{\Gamma_g^2 G^2 M^{2} m^{1/2}} \frac{1}{t^{7/2}}
	\\
	\zest \frac{M^{2} m^{1/6}}{[\Gcusp]^{1/3}} \frac{1}{t^{17/6}}
	\end{array} \right.
	+ (1 - f_{c}) \times {\rm Min} \left\{
	\begin{array}{l}
	\oet \frac{\Gkk}{\Gamma_g^{5/2} G^{5/2} M^{3} m} \frac{\log ( \Gamma_g G M^2 t / L_{\rm min} ) }{t^{4}} 
	\\
	\zsss \frac{M^{2} m^{1/4}}{[\Gkk]^{1/4}} \frac{1}{t^{11/4}}
	\end{array} \right. 
	\per
\end{align}
The integral $\int_{0}^{\infty} dL \, \nu_{g}(t,L) \, \PHHkk(L)$ is logarithmically sensitive to its lower limit.  
The divergence is artificial:  we should not be using $\nu_{g}$ as the distribution function for small loops with $L < L_{\rm min} \approx \Lkinky$, which decay predominantly by particle emission.  
In other words, the loop abundance is suppressed left of the black dots in \fref{fig:looplength}, but this suppression is not included in \eref{eq:PH_piecewise}.  
For numerical estimates, we will take the logarithmical factor to be a constant $O(10)$ number.  
By explicitly calculating \eref{eq:PH_def}, we have verified that \eref{eq:PH_approx} is an excellent approximation.  
Note also that the expressions in \eref{eq:PH_approx} become comparable in magnitude when $t \approx \tcuspy$ or $\tkinky$.  
As such, the minimization could be replaced with a step function on $t$ as in Refs.~\cite{Berezinsky:2011cp, Mota:2014uka}.

\section{Astrophysical and Cosmological Observables}\label{sec:Observables}

In this section we investigate how the string network may leave its imprint on astrophysical and cosmological observables.  
Since constraints associated with gravitational effects are universal (see, {\it e.g.}, \rref{Blanco-Pillado:2013qja} and references therein), we focus on constraints arising from the particle emission by the sting network.

\subsection{Big Bang Nucleosynthesis}\label{sub:BBN}

In the standard cosmological model, the abundances of the light elements were established by the process of Big Bang Nucleosynthesis (BBN) during the epoch $T_{\BBN} \approx (10-0.1) \MeV$ or $t_{\BBN} \approx (0.1-100) \sec$ (for a review see \rref{Olive:1999ij}).  
An injection of electromagnetic energy during this time can disrupt the remarkable agreement between the observed abundances and BBN's predictions, and therefore new physics can be highly constrained \cite{Sarkar:1995dd}.  
The emission of Higgs bosons from cusps and kinks on the dark string could be a potentially dangerous source of electromagnetic radiation.  
However, as we saw in \sref{sub:Friction} the network of dark strings remains friction dominated as long the the elastic Aharonov-Bohm scattering with the plasma is efficient, and during this time string loops are unable to radiate particles since cusps and kinks are smoothed by the friction.  
Friction domination terminates at a temperature $T_{\ast}$ given by \eref{eq:Tast}, which could be as late as the era of electron-positron annihilation at $T_{ann} \approx 0.1 \MeV$.  
One can consider two scenarios.  
In the first, the string tension is sufficiently low and the AB phase is sufficiently large that friction domination lasts all the way until $T_{ann} \approx 0.1 \MeV$.  
In this case, one does not expect the dark string model to be constrained at all from BBN since $0.1 \MeV$ corresponds to the end of BBN, and a particle production does not begin in earnest for a few Hubble times longer.  
In the second scenario, either the string tension is high or the AB phase is sufficiently small that friction domination ends well before the onset of BBN.  
It is this second case that we will focus on here.  

\begin{figure}[t]
\hspace{0pt}
\vspace{-0in}
\begin{center}
\includegraphics[width=0.49\textwidth]{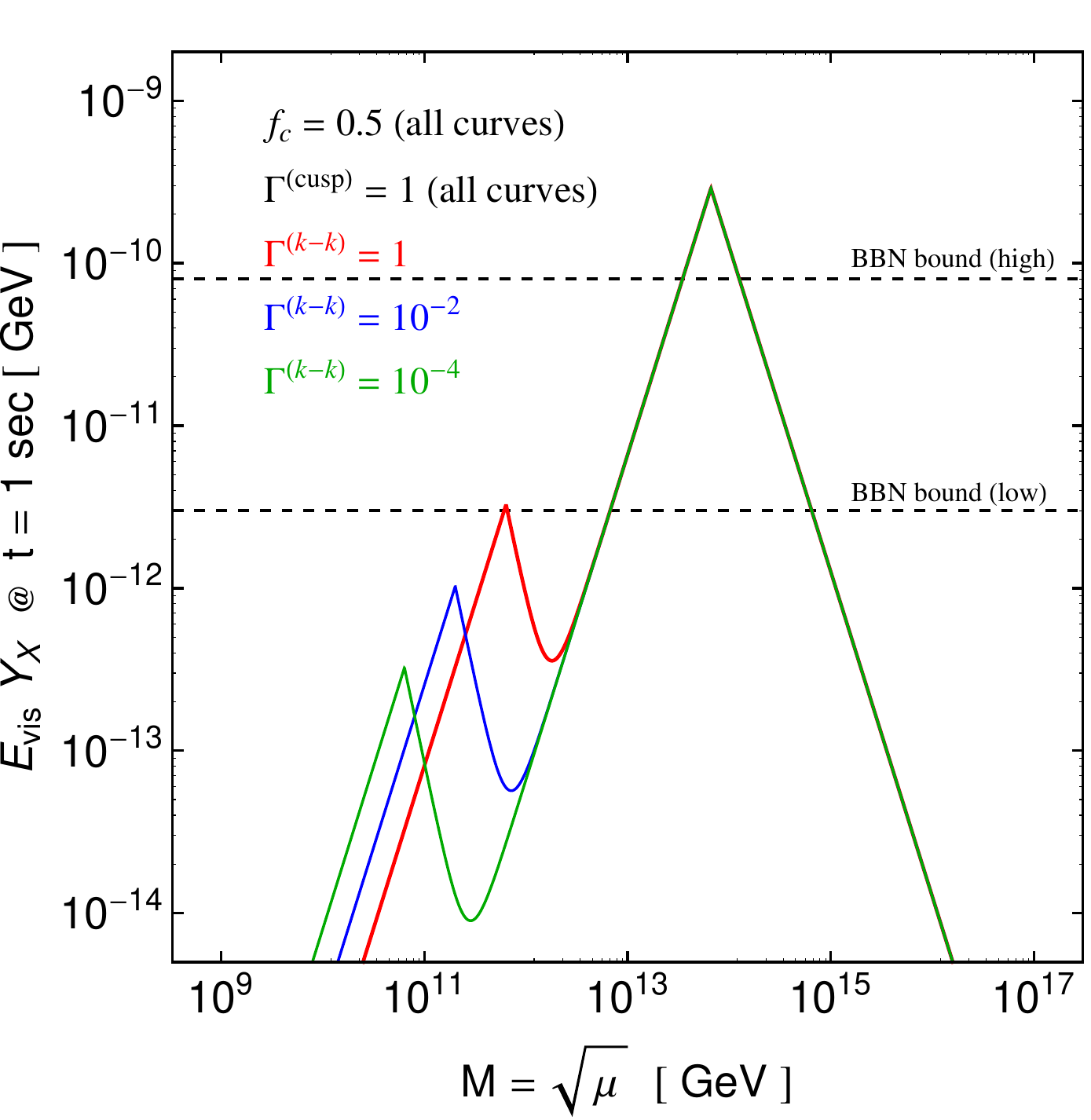} \hfill
\includegraphics[width=0.49\textwidth]{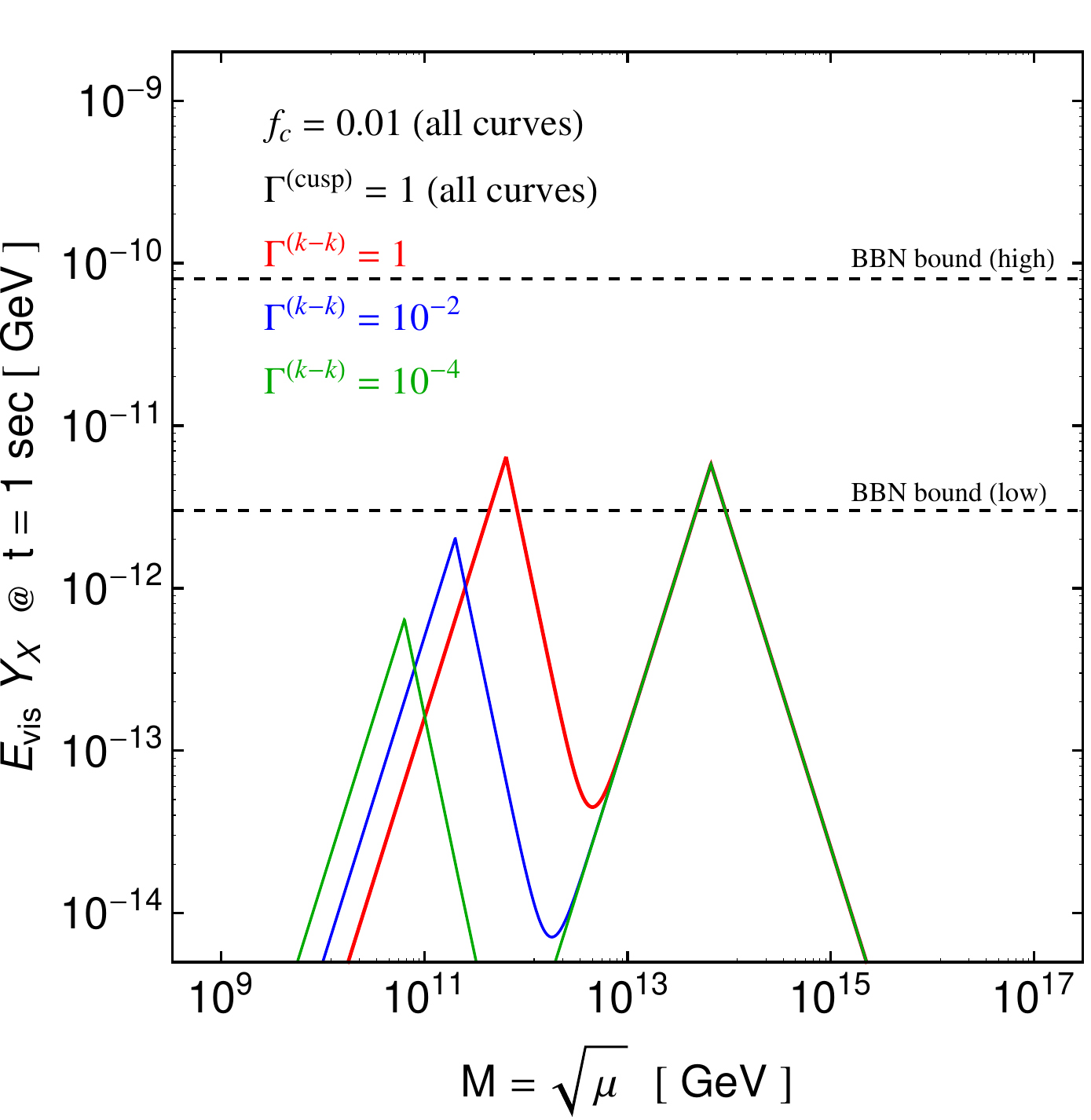} \\
\includegraphics[width=0.49\textwidth]{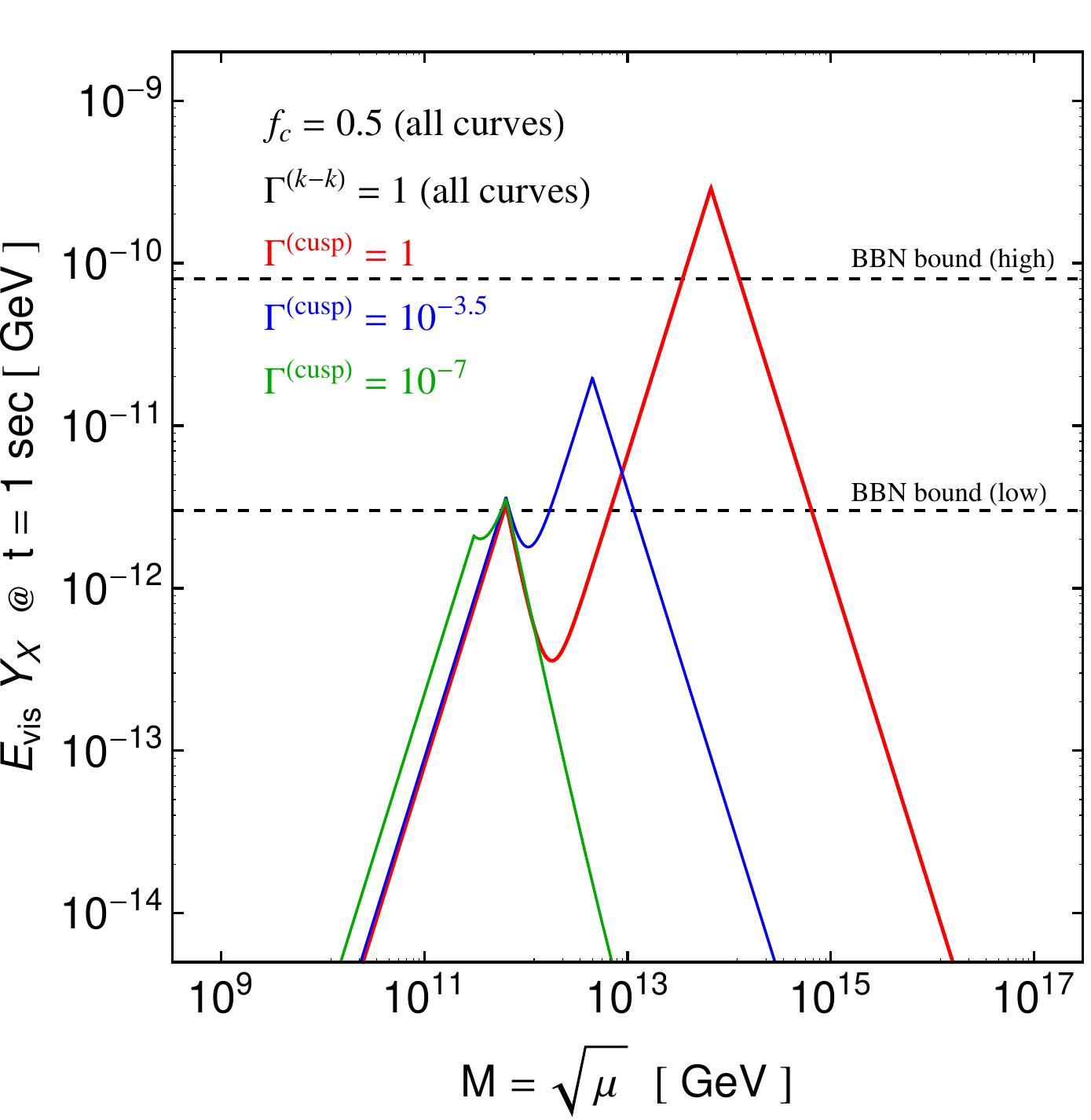} \hfill
\includegraphics[width=0.49\textwidth]{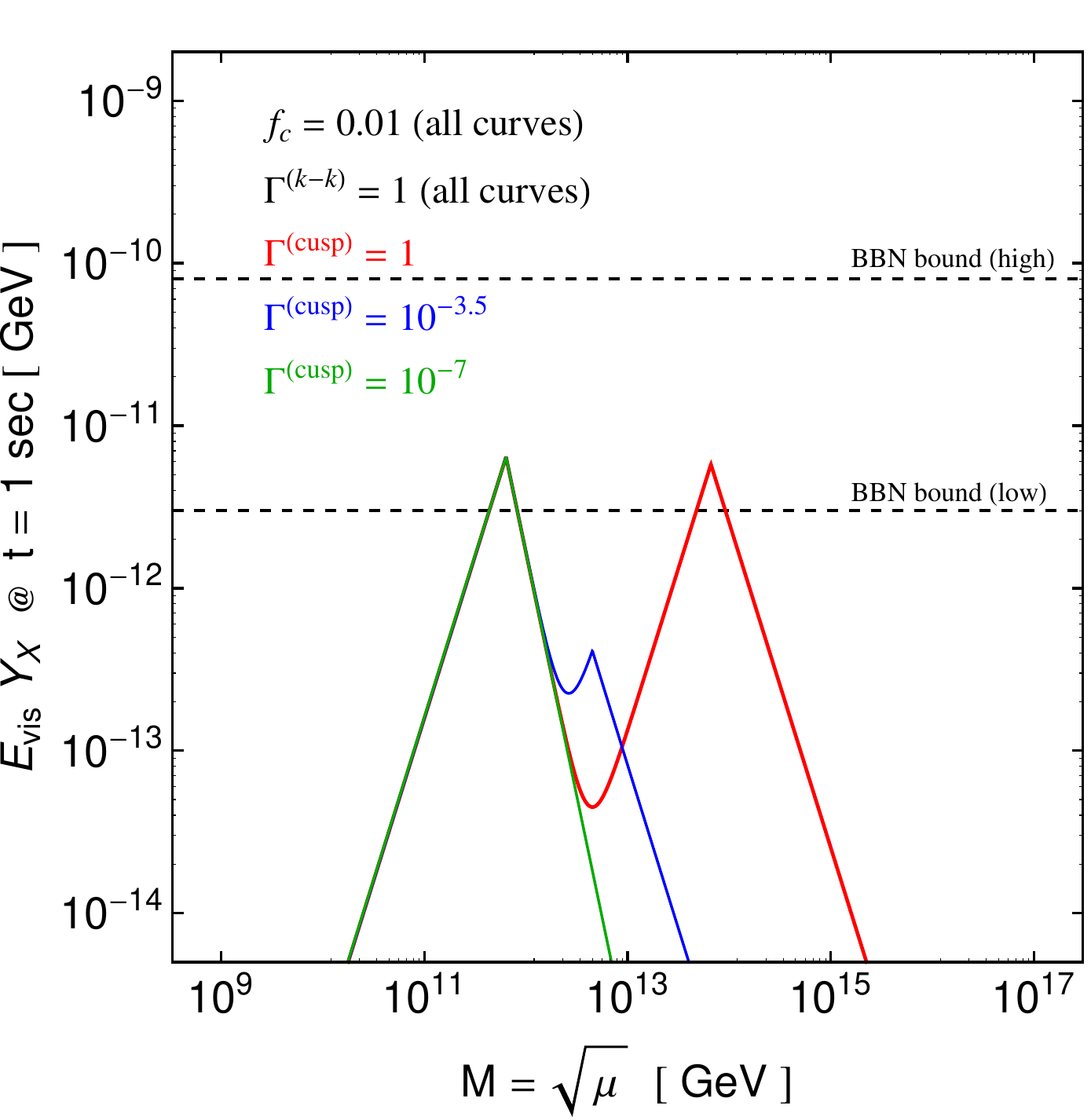} 
\caption{
\label{fig:EvisYX}
The predicted energy injection at $t = 1 \sec$, given by \eref{eq:EvisYX_approx}, compared to the bound from BBN in \eref{eq:BBN_bound}.  
The left panels show a cuspy loop network ($f_c = 0.5$), and the right panels shows a kinky loop network ($f_c = 0.01$).  
In the top panels, $\Gcusp = 1$ is fixed and $\Gkk$ is varied, and in the lower panels $\Gcusp$ is varied instead.
}
\end{center}
\end{figure}

\begin{figure}[t]
\hspace{0pt}
\vspace{-0in}
\begin{center}
\includegraphics[width=0.49\textwidth]{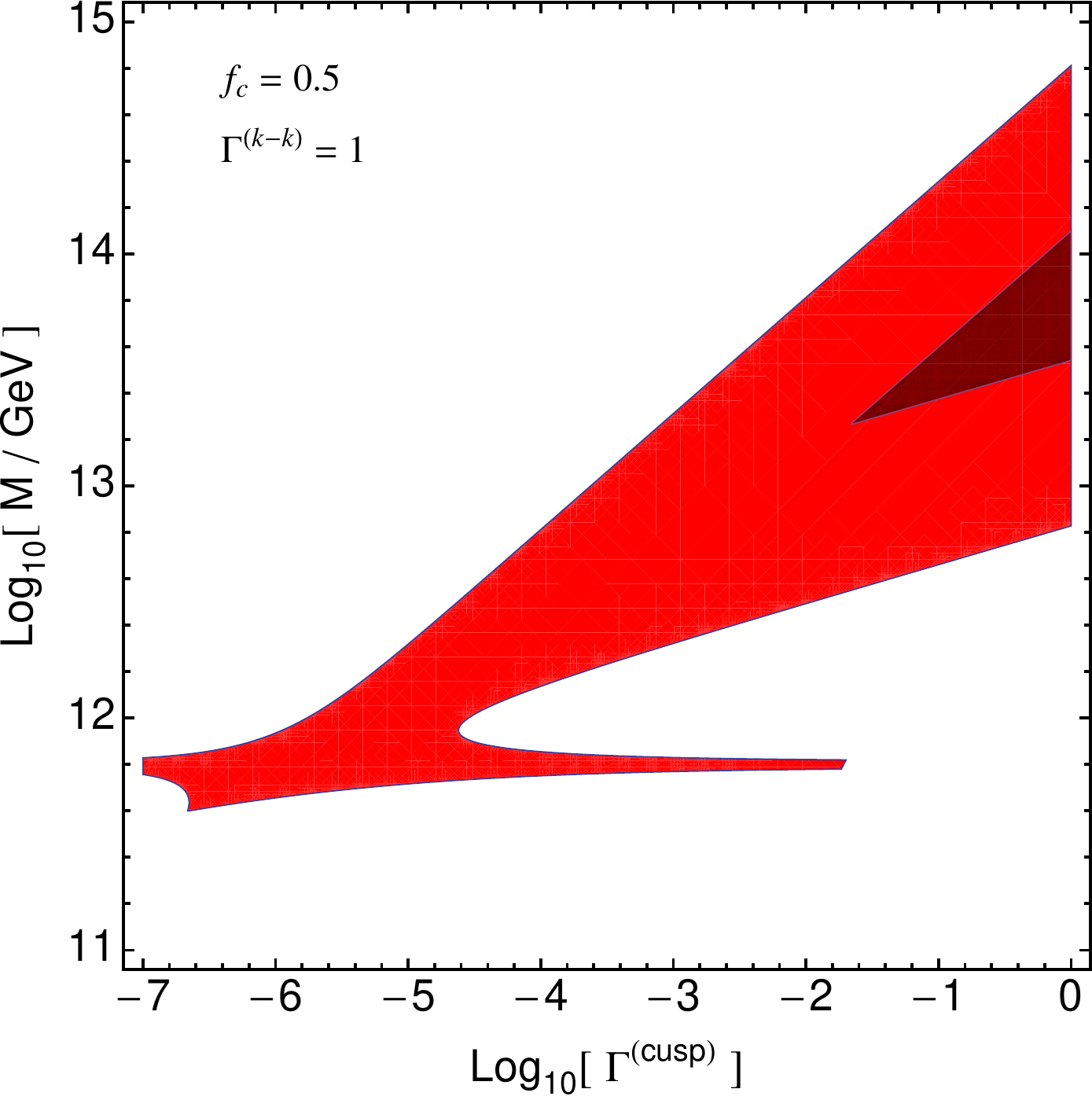} \hfill
\includegraphics[width=0.49\textwidth]{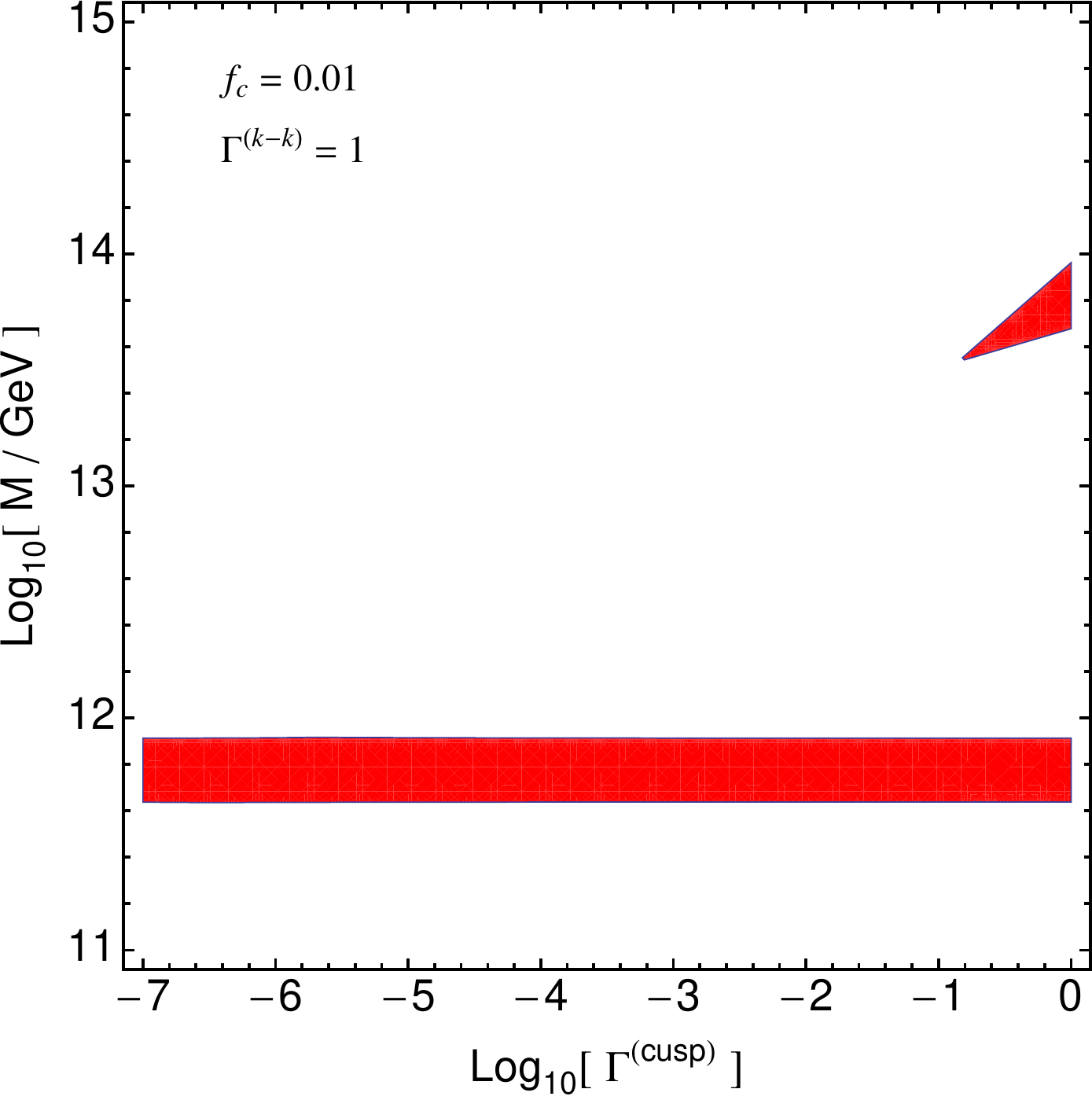} 
\caption{
\label{fig:BBN_bound}
The excluded region of parameter space for a cuspy string network (left) and a kinky string network (right).  
The contribution to $E_{\rm vis} Y_{X}$ from the kinky loops falls below the BBN bound for $\Gkk \lesssim 10^{-1}$, {\it i.e.} the region $M \approx 10^{12} \GeV$ is unconstrained if $\Gkk \lesssim 0.1$.  
}
\end{center}
\end{figure}

BBN constrains a network of cosmic strings \cite{Damour:1996pv, Peloso:2002rx} in much the same way that it constrains the late decay of a long-lived particle \cite{Cyburt:2002uv}.  
If a particle $X$ with number density $n_X(t)$ decays at time $10^{-1} \sec < t < 100 \sec$ and injects an energy $E_{\rm vis}$ in the form of (``visible'') electromagnetic radiation, then BBN constrains [see Fig. 38 of \rref{Kawasaki:2004qu}] 
\begin{align}\label{eq:BBN_bound}
	E_{\rm vis} Y_{X}(t \approx 1 \sec) \lesssim  (3 - 80) \times 10^{-12} \GeV
\end{align}
where $Y_X = n_X / s$ is the dimensionless yield and $s(T) = (2\pi^2 g_{\ast S}/45) T^3$ is the entropy density with $g_{\ast S} \approx 10$ during BBN.  
During the radiation era we have $T = T_{eq} (t_{eq} / t)^{1/2}$.  
The range in \eref{eq:BBN_bound} corresponds to different measurements of the primordial hydrogen abundance, and we refer to these as the ``low'' and ``high'' bounds in our analysis.
The bound depends on the nature of the decay products, which are primarily hadronic in the case of Higgs decays.  
This corresponds to a more stringent bound than leptonic decays.  

For the dark string network, we calculate the energy injection as 
\begin{align}
	E_{\rm vis} Y_{X} = t \frac{\Pcal_{H}}{s}
\end{align}
where $\Pcal_{H}$ is given by \eref{eq:PH_def}.  
Making use of the approximation in \eref{eq:PH_approx} we find
\begin{align}\label{eq:EvisYX_approx}
	E_{\rm vis} Y_{X} \approx & \ 
	\frac{f_{c}}{t_{eq}^{3/2}T_{eq}^3} \times {\rm Min} \left\{
	\begin{array}{l}
	\zffn \frac{\Gcusp}{\Gamma_g^2 G^2 M^{2} m^{1/2}} \frac{1}{t} 
	\\
	\zons \frac{M^{2} m^{1/6}}{[\Gcusp]^{1/3}} \frac{1}{t^{1/3}} 
	\end{array} \right.
	+ \frac{(1 - f_{c})}{t_{eq}^{3/2}T_{eq}^3} \times {\rm Min} \left\{
	\begin{array}{l}
	\fos \frac{\Gkk}{\Gamma_g^{5/2} G^{5/2} M^{3} m} \frac{1}{t^{3/2}} 
	\\
	\zoff \frac{M^{2} m^{1/4}}{[\Gkk]^{1/4}} \frac{1}{t^{1/4}} 
	\end{array} \right. 
\end{align}
In \fref{fig:EvisYX} we plot $E_{\rm vis} Y_{X}$ evaluated at $t = 1 \sec$ as a function of the string mass scale $M = \sqrt{\mu}$.  
In the case of a cuspy loop network, $f_c \approx 0.5$, we find that the BBN bound is exceeded if the string mass scale falls in the range $10^{12} \GeV \lesssim M \lesssim 10^{15} \GeV$ and if $\Gcusp$ is sufficiently large.
In the case of a kinky loop network, $f_c \approx 0.01$, the bound is only exceeded for a very narrow range of parameters.  
(Recall that $\Gcusp, \Gkk < O(1)$ for realistic models as in \erefs{eq:Gamma_HH}{eq:Gamma_H}.) 
The double-peaked structure is a consequence of summing the two populations of cuspy and kinky loops.  
If the string mass scale is high (low), $E_{\rm vis} Y_X$ is dominated by the contribution from cuspy loops (kinky loops).  
The peaks occur at the mass scales 
\begin{align}
	M_{\rm cuspy}( t = 1 \sec ) & 
	\simeq ( 5 \times 10^{13} \GeV ) \left( \Gcusp \right)^{1/3} 
	\\
	M_{\rm kinky} ( t = 1 \sec ) &
	\simeq  ( 3 \times 10^{11} \GeV ) \left( \Gkk \right)^{1/4} 
\end{align}
where $M_{\rm cuspy}(t)$ and $M_{\rm kinky}(t)$ are given by \eref{eq:M_cuspy_kinky}.  %

The excluded parameter regime is also shown in \fref{fig:BBN_bound}.  
The bound arising from cuspy loops agrees well with the recent calculation of \rref{Mota:2014uka}.  
Additionally, there is a bound from kinky loops if $M = (6- 9) \times 10^{11} \GeV$ and $\Gkk \approx 1$.  
For $\Gkk \lesssim 0.1$ the predicted energy injection does not exceed the BBN bound, see the top-right panel of \fref{fig:EvisYX}.  

\subsection{CMB Spectral Distortion}\label{sub:SpectralDist}

For the most part, Higgs emission during the radiation era leaves no imprint, since the Higgs decays quickly into particles that thermalize with the plasma.  
Late into the radiation era, however, thermalization is inefficient and the energy injection can manifest itself as a distortion of the CMB spectrum \cite{Zeldovich:1969ff, Sunyaev:1970er} (see also the review \cite{Tashiro:2014pga}).  
Specifically, if the radiation is injected after the photon-number-violating double Compton scattering process has gone out of equilibrium, the spectrum will be of the Bose-Einstein form with a nonzero dimensionless chemical potential $\mudist$.
The magnitude of the so-called $\mu$-distortion is constrained by the COBE FIRAS measurement of the CMB spectrum as $|\mudist| < \mu_{\FIRAS}$ with \cite{Fixsen:1996nj}
\begin{align}\label{eq:mu_FIRAS}
	\mu_{\FIRAS} = 9 \times 10^{-5} 
\end{align}
at the $95\%$ confidence level.  
The next-generation CMB telescope PIXIE is expected to achieve a sensitivity of \cite{Kogut:2011xw}
\begin{align}\label{eq:mu_PIXIE}
	\delta \mu_{\PIXIE} \approx 5 \times 10^{-8} \per 
\end{align}

The chemical potential $\mudist$ measures the ratio of the injected energy, $\Delta \rho_{\gamma}$, to the energy of the blackbody spectrum, $\rho_{\gamma} = (2 \pi^2 / 30) T^4$.  
The $\mu$-distortion is induced when the energy injection occurs after double Compton scattering goes out of equilibrium ($t_1 \approx 1 \yr$, $z_1 \approx 2 \times 10^{6}$) and before single Compton scattering goes out of equilibrium ($t_2 \approx 100 \yr$, $z_2 \approx 10^{5}$).  
Earlier energy injections can still thermalize, and later energy injections lead to a $y$-type distortion, but we do not expect this to provide a stronger constraint (see, {\it e.g.} \rref{Tashiro:2012nb}).  
The rate of energy density injection in the form of Higgs bosons was given by $\Pcal_{H}(t)$ in \eref{eq:PH_def}, and we can estimate the contribution that goes into electromagnetic species with a factor of $f_{em} \approx 1/2$ since roughly half of the energy is lost into neutrinos \cite{Berezinsky:1998ft}.  
The $\mu$-distortion is estimated as \cite{Tashiro:2014pga} 
\begin{align}\label{eq:mu_def}
	\mudist \approx 1.4 \frac{\Delta \rho_{\gamma}}{\rho_{\gamma}} = 1.4 f_{em} \int_{t_1}^{t_2} dt \, \frac{\Pcal_{H}(t)}{\rho_{\gamma}(t)} \per
\end{align}
During the radiation era we have $(1 + z(t)) / (1 + z_{eq})= (t_{eq} / t)^{1/2} = T(t) / T_{eq}$.  
This estimate neglects the evolution of $\mudist(t)$ until recombination ($z_{rec} \approx 1300$) but it is reliable for order of magnitude estimates \cite{Chluba:2011hw}.  

\begin{figure}[t]
\hspace{0pt}
\vspace{-0in}
\begin{center}
\includegraphics[width=0.49\textwidth]{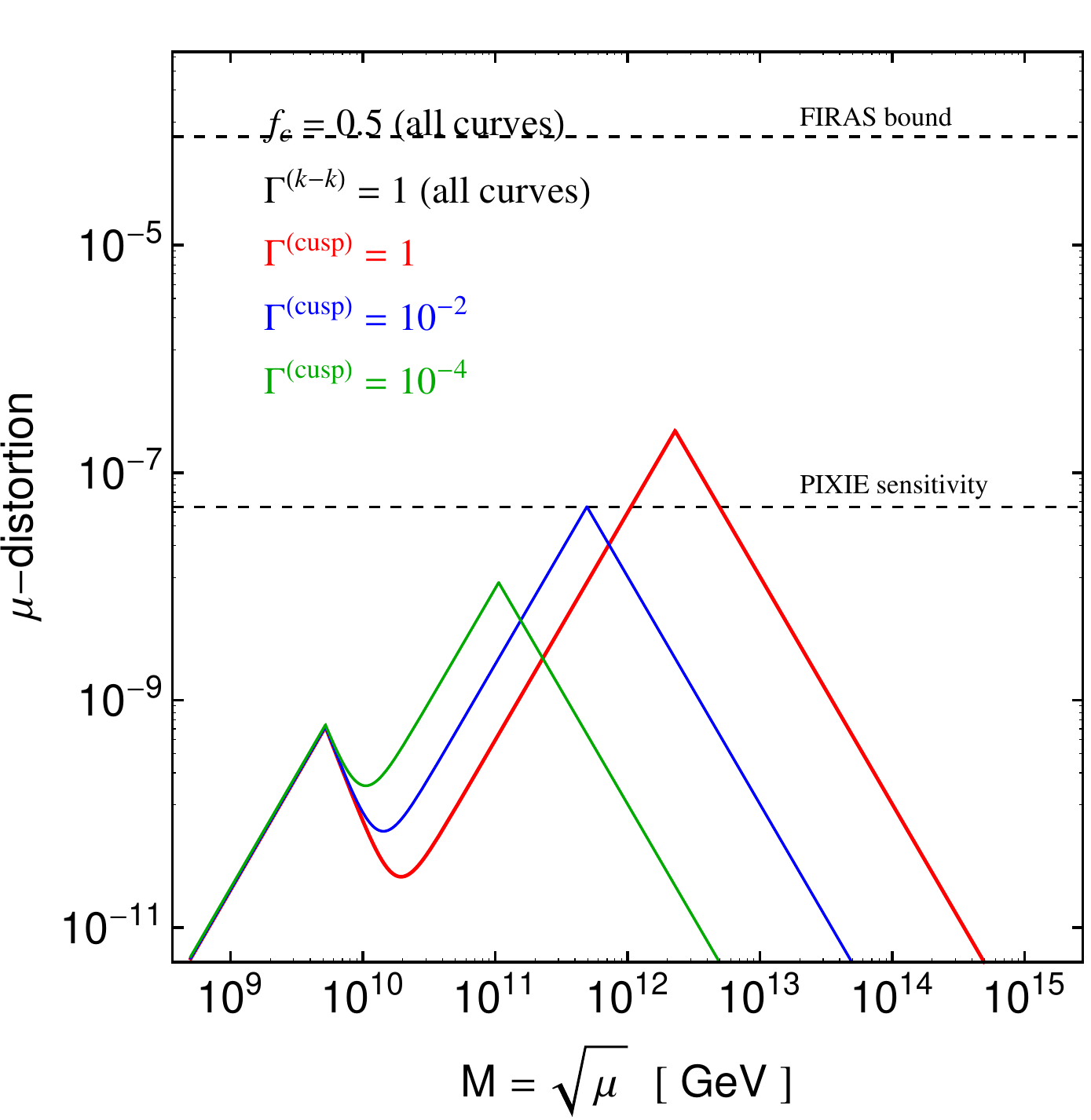} \hfill 
\includegraphics[width=0.49\textwidth]{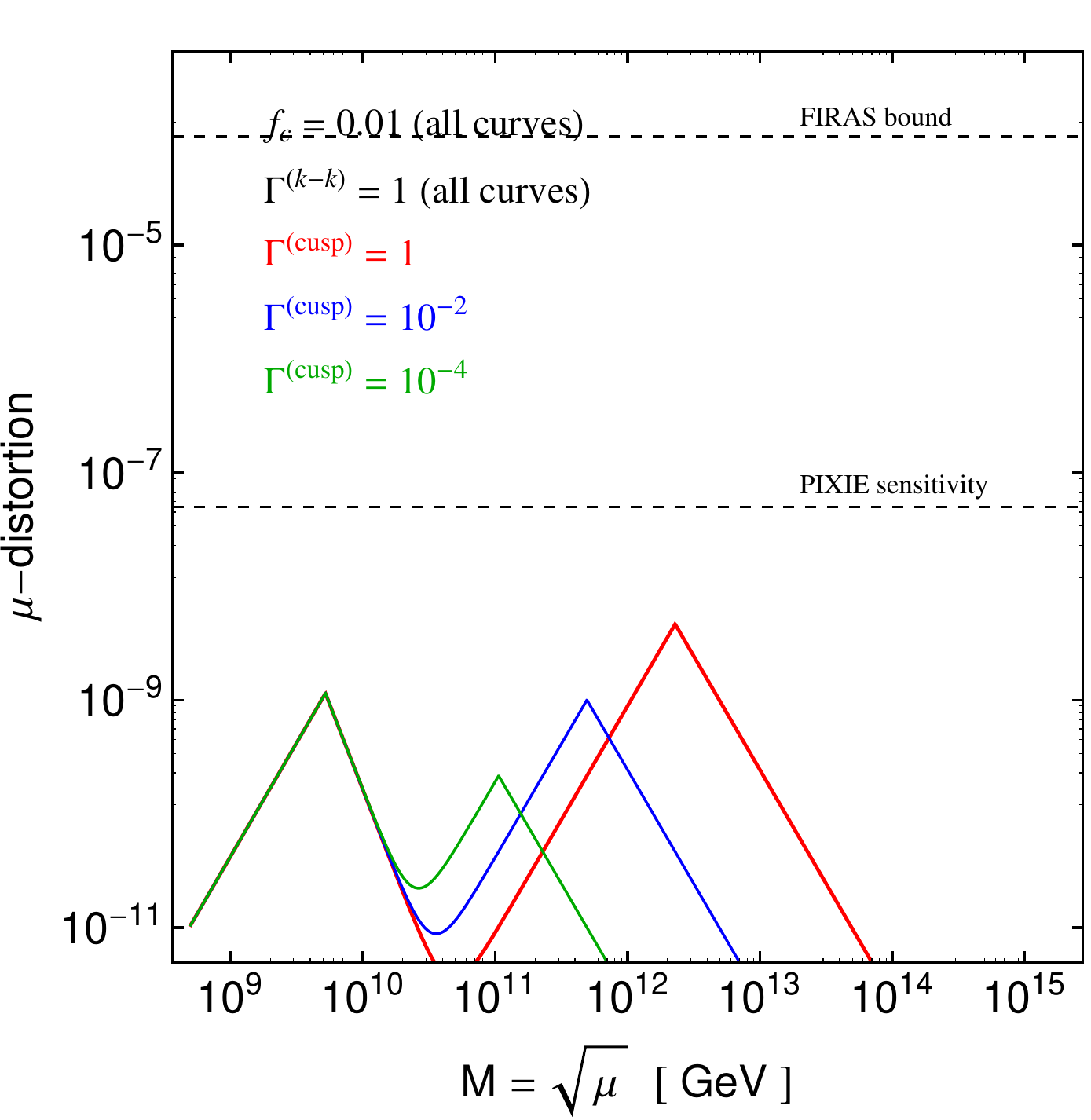} 
\caption{
\label{fig:mu_dist}
The predicted $\mu$-distortion, given by \eref{eq:mu_approx}, compared to the FIRAS bound and expected sensitivity of PIXIE.  
Since all curves fall well below the FIRAS limit, $\mu_{\FIRAS} = 9 \times 10^{-5}$, the model is currently unconstrained.  
}
\end{center}
\end{figure}

We estimate $\mudist$ using the the piecewise approximation to $\Pcal_H(t)$ given in \eref{eq:PH_approx}.  
Then we find 
\begin{align}\label{eq:mu_approx}
	\mudist \approx \ & 
	\frac{f_{em} f_{c}}{T_{eq}^4 t_{eq}^2} \times {\rm Min} \left\{ \begin{array}{l}
	\sff \frac{\Gcusp}{\Gamma_g^2 G^2 M^{2} m^{1/2}} \frac{1}{t_1^{1/2}} \\
	\ses \frac{M^{2} m^{1/6}}{[\Gcusp]^{1/3}} t_{2}^{1/6}
	\end{array} \right. 
	+
	\frac{f_{em} ( 1 - f_{c} )}{T_{eq}^4 t_{eq}^2} \times {\rm Min} \left\{ \begin{array}{l}
	\tsn \frac{\Gkk}{\Gamma_g^{5/2} G^{5/2} M^{3} m} \frac{1}{t_1} \\
	\foo \frac{M^{2} m^{1/4}}{[\Gkk]^{1/4}} t_{2}^{1/4}
	\end{array} \right. 
\end{align}
\fref{fig:mu_dist} shows the predicted $\mu$-distortion for a range of model parameters: varying Higgs-to-string coupling $\gHHstr$ and varying string mass scale $M$.  
The peaks occur at the mass scales 
\begin{align}
	M_{\rm cuspy}( t_{1}^{3/4} t_{2}^{1/4} ) & 
	\simeq ( 2 \times 10^{12} \GeV ) \left( \Gcusp \right)^{1/3}
	\\
	M_{\rm kinky} ( t_{1}^{4/5} t_{2}^{1/5} ) &
	\simeq  ( 3 \times 10^{9} \GeV ) \left( \Gkk \right)^{1/4}
\end{align}
where $M_{\rm cuspy}(t)$ and $M_{\rm kinky}(t)$ are given by \eref{eq:M_cuspy_kinky}.  

For the entire parameter range indicated, the expected level of spectral distortion is well-below the current FIRAS bound, \eref{eq:mu_FIRAS}, and the model is unconstrained.  
A future experiment with a sensitivity comparable to PIXIE has the potential to constrain $M \approx 10^{12}-10^{13} \GeV$ if $\Gcusp \gtrsim 10^{-2}$.

\subsection{Diffuse Gamma Ray Background}\label{sub:DiffuseGamma}

As we saw in \sref{sec:SM_Particle_Prod}, particle emission from the dark string is mostly in the form of SM Higgs bosons.  
The Higgs decays very quickly, mostly into hadrons such as pions, which themselves decay into photons and charged leptons \cite{Djouadi:2005gi}.  
These particles can scatter on the CMB photons and extragalactic background light initiating an electromagnetic (EM) cascade \cite{Ginzburg:1990sk}.  
This process results in gamma rays that will show up on Earth as a diffuse gamma ray background (DGRB).  

The Fermi-LAT gamma ray telescope measures the spectrum of the DGRB in the energy range $100 \MeV \lesssim E \lesssim 100 \GeV$, and the spectrum is seen to fall as $E^{-2.4}$ \cite{Abdo:2010nz}.  
If the DGRB arose from an EM cascade, then the spectrum would be softer, falling as $E^{-2}$ \cite{Ginzburg:1990sk}.  
The cascade photons, therefore, can only make up a subdominant contribution to the total DGRB, implying an upper bound on the amplitude of this component of the spectrum.  
Since the predicted spectral shape of the EM cascade is known, it is convenient to express this limit instead as a bound on the integrated spectrum, {\it i.e.} the total energy density $\omega_{\rm cas}$ in the EM cascade today \cite{Berezinsky:1998ft}.  
This bound is found to be $\omega_{\rm cas} < \omega_{\rm cas}^{\rm max}$ with \cite{Berezinsky:2010xa}
\begin{align}\label{eq:omegacas_max}
	\omega_{\rm cas}^{\rm max} \approx 5.8 \times 10^{-7} \frac{\eV}{\cm^3} \per
\end{align}


We calculate $\omega_{\rm cas}$ using the Higgs injection function $\Pcal_{H}(t)$ given by \eref{eq:PH_def}.  
Not all of the energy carried by the Higgs bosons is transferred to the EM cascade.  
We estimate the fractional contribution as $f_{em} \approx 1/2$ since roughly half of the energy is lost into neutrinos \cite{Berezinsky:1998ft}.  
Then $\omega_{\rm cas}$ is calculated by integrating over the history of the network as \cite{Ginzburg:1990sk} (see also \rref{Berezinsky:1998ft})
\begin{align}\label{eq:omegacas_def}
	\omega_{\rm cas} \equiv f_{em} \int_{t_{\rm cas}}^{t_{\rm today}} dt \, \Pcal_H(t) \frac{1}{[1+z(t)]^4} 
\end{align}
where the factor of $(1 + z)^{-4} = (t / t_{\rm today})^{8/3}$ accounts for the cosmological redshift between time $t$ in the matter era and today ($t = t_{\rm today}$).  
We truncate the integral at $t_{\rm cas} \approx 10^{15} \sec$ (or $z_{\rm cas} \approx 60$) corresponding approximately to the time at which the universe became transparent to gamma rays \cite{Berezinsky:2011cp}.  

We perform the integrals in \eref{eq:omegacas_def} using the approximation to the Higgs injection function, given by \eref{eq:PH_approx}, and we obtain the predicted EM cascade energy density today to be 
\begin{align}\label{eq:omegacas_approx}
	\omega_{\rm cas} 
	= \ &  
	f_{em} \, f_{c} \times 
	{\rm Min}
	\begin{cases}
	\stf \ 
	\frac{\Gcusp}{\Gamma_g^2 G^2 M^{2} m^{1/2}} \ 
	\frac{t_{eq}^{1/2} }{t_{\rm today}^{8/3} t_{\rm cas}^{1/3} }
	& 
	\\
	\tfn \ 
	\frac{M^{2} m^{1/6}}{[ \Gcusp]^{1/3}} \ 
	\frac{t_{eq}^{1/2}}{t_{\rm today}^{7/3}} 
	& 
	\end{cases} 
	\nn & 
	+ f_{em} \, (1-f_c) \times 
	{\rm Min} 
	\begin{cases}
	\tto \ 
	\frac{\Gkk}{\Gamma_g^{5/2} G^{5/2} M^{3} m} \ 
	\frac{t_{eq}^{1/2}}{t_{\rm today}^{8/3} t_{\rm cas}^{5/6}}
	& 
	\\
	\ost \ 
	\frac{M^{2} m^{1/4}}{[ \Gkk]^{1/4}} \ 
	\frac{t_{eq}^{1/2}}{t_{\rm today}^{9/4}}
	& 
	\end{cases}
\end{align}
where we have used $t_{\rm cas} \ll t_{\rm today}$.  
Note that for the case in which gravitational (particle) radiation controls the loop decay, the integral is dominated by the lower (upper) limit of integration corresponding to early (late) times.  
This behavior results from a competition between the Higgs injection function $\Pcal_H$, which decreases with increasing $t$, and the redshift factor, which increases with increasing $t$.  

The contribution to $\omega_{\rm cas}$ from cuspy loops has been calculated previously in Refs.~\cite{Dufaux:2012np, Berezinsky:2011cp, Mota:2014uka}, and our result (first term of \eref{eq:omegacas_approx}) agrees with those references.
\rref{Vachaspati:2009kq} neglects the radiation era relic loops and obtains a different expression for $\omega_{\rm cas}$.  
The contribution to $\omega_{\rm cas}$ from kinky loops has been calculated in \rref{Lunardini:2012ct}, but this cannot be compared against our \eref{eq:omegacas_approx} (second term) since \rref{Lunardini:2012ct} takes the emission power from kinks to be of the form $P \sim L^{0}$ whereas we have $P \sim L^{-1}$ ({\it c.f.} \eref{eq:P_kk}).  

We are now led to ask why our calculation matches well with the literature even though the prior work has overlooked the factor of $(L/L_0)^{\idx}$, which was discussed below \eref{eq:loop_dist}.  
Since this factor suppresses the abundance of small loops, $L \leq L_0$, we only expect it to have an impact when small loops give the dominant contribution to $\Pcal_H$.  
Three of the four integrals in \eref{eq:PH_piecewise} are insensitive to the lower limit of integration, and the fourth integral has only a logarithmic sensitivity, discussed below \eref{eq:PH_approx}.  
If the particle radiation power would grow more rapidly with decreasing $L$, then a more dramatic effect would be seen upon implementing the correct loop length distribution.  

\begin{figure}[t]
\hspace{0pt}
\vspace{-0in}
\begin{center}
\includegraphics[width=0.49\textwidth]{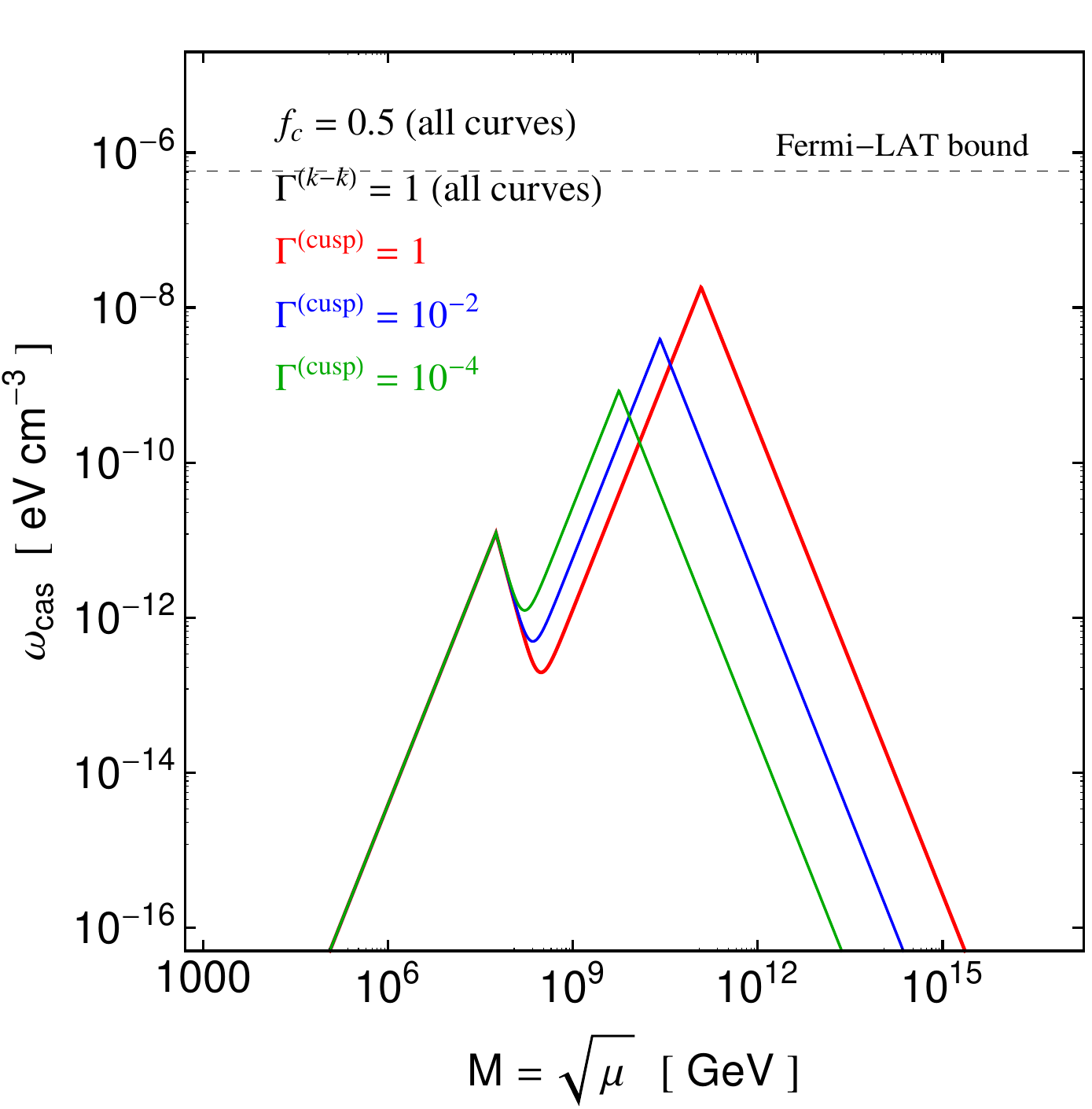} \hfill 
\includegraphics[width=0.49\textwidth]{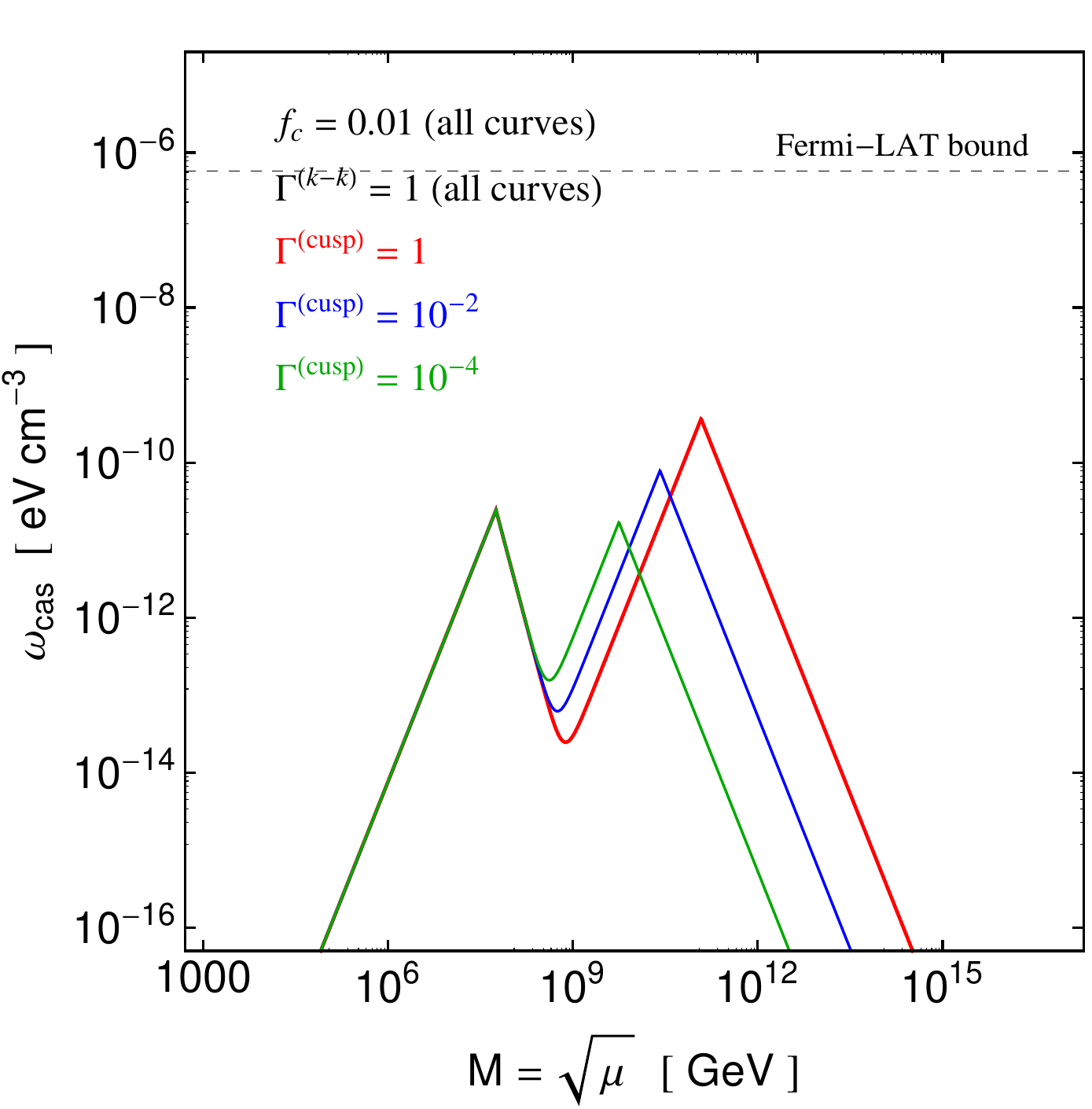} 
\caption{
\label{fig:wcas}
The EM cascade energy density, given by \eref{eq:omegacas_approx}, compared to the Fermi-LAT upper bound, given by \eref{eq:omegacas_max}.  
}
\end{center}
\end{figure}

In \fref{fig:wcas} we show $\omega_{\rm cas}$ for a range of values of the string tension scale $M = \sqrt{\mu}$.  
Even for the maximum reasonable value of the dimensionless couplings, $\Gcusp \sim \Gkk \sim 1$ we find that the predicted $\omega_{\rm cas}$ falls below the Fermi-LAT bound, and the model is unconstrained.  
This conclusion was also obtained by \rref{Mota:2014uka}.
The peaks occur at the mass scales
\begin{align}
	M_{\rm cuspy}( t_{\rm today}^{1/2} t_{\rm cas}^{1/2} ) & 
	\simeq ( 1 \times 10^{11} \GeV ) \left( \Gcusp \right)^{1/4}
	\\
	M_{\rm kinky} ( t_{\rm today}^{1/3} t_{\rm cas}^{2/3} ) &
	\simeq  ( 4 \times 10^{7} \GeV ) \left( \Gkk \right)^{1/3}
\end{align}
where $M_{\rm cuspy}(t)$ and $M_{\rm kinky}(t)$ are given by \eref{eq:M_cuspy_kinky}.  
For large (small) $M$ the EM cascade primarily arises from the cuspy (kinky) loops.  
Eventhough the rate of Higgs emission from kinky loops is smaller than cuspy loops by a factor of $1 / \sqrt{ML}$ (compare \erefs{eq:P_cusp}{eq:P_kk}), these loops dominate $\omega_{\rm cas}$ since they decay more slowly than the cuspy loops and therefore have a larger abundance today (see \sref{sub:LoopDensity}).

\subsection{Diffuse Cosmic Ray Fluxes}\label{sub:DiffuseCR}

Since the 1980s there has been considerable interest in the study of high energy cosmic rays (CRs) produced by cosmic string networks (see the review \cite{Berezinsky:1998ft}).  
For the case of non-superconducting strings the conclusions of these studies were often negative, meaning that the predicted CR flux fell short of the observed flux \cite{Srednicki:1986xg, MacGibbon:1989kk, Bhattacharjee:1989vu, Gill:1994ic} (see also \cite{Hill:1986mn, Ostriker:1986xc}).  
Since the CR calculation parallels our discussion of the DGRB in \sref{sub:DiffuseGamma}, and since the latter bound is typically more constraining \cite{Berezinsky:2010xa}, we will only briefly review the diffuse CR proton and neutrino fluxes.  

{\bf (i) Protons} \\
The Higgs decay leads to a hadronic cascade that can produce cosmic ray protons.  
In 1989 \rref{Bhattacharjee:1989vu} was the first to study the CR proton flux that results when $X$ bosons are emitted in cusp evaporation events and subsequently decay hadronically.  
This analysis concluded that the model is unconstrained regardless of the string tension mass scale.  
The results of this reference to not directly carry over to the case of the dark string, because \rref{Bhattacharjee:1989vu} took the particle radiation power to scale as $P \propto L^{-1/3}$ whereas we have $L^{-1/2}$ (see \eref{eq:P_cusp}).  
However, the general conclusion is the same, which we will demonstrate with the following estimate.  

The observed flux of ultra-high energy cosmic ray protons is measured to be \cite{Abbasi:2007sv, Abraham:2010mj, AbuZayyad:2012ru}
\begin{align}\label{eq:proton_flux_obs}
	E^3 I_{p} \bigr|_{obs.} \approx 10^{24} \eV^2 \, {\rm m}^{-2} \sec^{-1} \sr^{-1}
\end{align}
for $2 \times 10^{17} \eV < E < 5 \times 10^{19} \eV$.  
We suppose that a fraction $f_p$ of the energy emitted from the cosmic string network is transferred to CR protons.  
We can estimate this energy density, call it $\omega_{\CR}$, from the EM cascade energy density given previously by \eref{eq:omegacas_def} using $\omega_{\CR} \approx (f_p / f_{em}) \omega_{\rm cas}$.  
From \fref{fig:wcas} we see that in the most optimistic region of parameter space $\omega_{\CR} \approx f_{p} \times 10^{-8} \eV \cm^{-3}$.  
At energies of $10^{19} \eV$ this roughly corresponds to a flux of 
\begin{align}\label{eq:proton_flux}
	E^3 I_{p} \approx \frac{E}{4\pi} \omega_{\CR} \approx \bigl( 10^{23} \eV^2 \, {\rm m}^{-2} \sec^{-1} \sr^{-1} \bigr) \frac{f_p}{0.1} \per
\end{align}
We find that the predicted flux is insufficient to explain the observed flux in \eref{eq:proton_flux_obs}.  

{\bf (ii)  Ultra-High Energy Neutrinos} \\
Ultra-high energy neutrinos will typically be produced by the same decay chain responsible for the electromagnetic cascade discussed in \sref{sub:DiffuseGamma}.  
Namely, the Higgs boson decay initiates a hadronic cascade, and neutrinos are produced along with gamma rays from the pion / kaon decays.  
Consequently, the predicted neutrino flux is tied to the flux of the EM cascade gamma rays \cite{Berezinsky:1975zz} (see also \rref{Ginzburg:1990sk}).  
If the neutrino spectrum scales as $J_{\nu} \propto E^{-2}$ over the range $E_{\rm min} < E < E_{\rm max}$ then the amplitude is bounded as
 \cite{Berezinsky:2005rw} 
\begin{align}\label{eq:nuflux_to_omegacas}
	E^2 \, J_{\nu} \leq \frac{1}{4\pi} \frac{\omega_{\rm cas}}{\ln E_{\rm max} / E_{\rm min} }
\end{align}
where $\omega_{\rm cas}$ is the energy density of the cascade photons today (see \eref{eq:omegacas_def}).  
As we saw in \fref{fig:wcas}, the most optimistic prediction of the dark string model is $\omega_{\rm cas} \approx 10^{-8} \eV \cm^{-3}$.  
Then estimating the logarithm as $O(1)$ we find the most optimistic prediction of the dark string model to be 
\begin{align}\label{eq:nuflux_max}
	E^2 \, J_{\nu} \approx 2 \times 10^{-8} \GeV \cm^{-2} \sec^{-1} \sr^{-1} \per
\end{align}
This is still well below the current best limit from Ice Cube \cite{Aartsen:2013dsm}
\begin{align}\label{eq:IceCube_bound}
	E^2 \, J_{\nu} \bigr|_{\IceCube} \lesssim 1.2 \times 10^{-7} \GeV \cm^{-2} \sec^{-1} \sr^{-1}
\end{align}
for $10^{6} \GeV \lesssim E \lesssim 10^{11} \GeV$.  
Future experiments such as LOFAR and SKA expect sensitivities at the level of $10^{-9}$ and $10^{-10} \GeV \cm^{-2} \sec^{-1} \sr^{-1}$ \cite{James:2008ff}.  
This sensitivity may be sufficient to detect a flux at the level given by \eref{eq:nuflux_max}, but since \eref{eq:nuflux_to_omegacas} is an upper bound, a more careful calculation of $J_{\nu}$ is required to assess whether these experiments will be able to probe the model.

\subsection{Burst Rate}\label{sub:BurstRate}

When radiation is emitted from a cusp on a cosmic string, it is beamed into a narrow cone \cite{Vachaspati:1984gt}.  
If this radiation reaches the Earth, it may be observed as a burst instead of a diffuse flux \cite{Damour:2001bk, Berezinsky:2009xf}.  
Let $\Gamma_{\burst}(t) dt$ be the number of bursts that arrive at the Earth between time $t$ and $t+dt$.  
We will estimate $\Gamma_{\burst}(t_{\rm today})$ assuming that (i) every burst directed toward the Earth eventually reaches the Earth (no dimming) and (ii) the radiation from a burst propagates in a straight line at the speed of light.  
Then, $\Gamma_{\burst}(t)$ is obtained by counting the number of loops in the past light cone of the Earth at time $t$, multiplying by the burst rate, and including a geometrical factor to account for the burst that are not directed toward the Earth.  

Let $R(t,t^{\prime})$ be the radius at time $t^{\prime}$ of the past lightcone of an event at time $t > t^{\prime}$.  
During the matter era, it is given by $R(t,t^{\prime}) = a(t^{\prime}) \int_{t^{\prime}}^{t} dt^{\prime \prime} / a(t^{\prime \prime}) = 3 t^{1/3} (t^{\prime})^{2/3} [ 1 - (t^{\prime} / t)^{1/3} ]$.  
Then the time interval from $t$ to $t+dt$ has a past light cone (a conical shell) that fills a spatial volume at time $t^{\prime}$ given by\footnote{In redshift space with $t=t_{\rm today}$ this becomes $dV(z^{\prime}) = 54 \pi t_{\rm today}^3 (\sqrt{1+z^{\prime}} -1)^2 (1+z^{\prime})^{-5} dz^{\prime}$.  For comparison, the volume contained between two surfaces of fixed redshift is smaller: $d\bar{V}(z^{\prime}) = dV(z^{\prime}) (1+z^{\prime})^{-1/2}$ \cite{Mukhanov:2005}.  
}
\begin{align}\label{eq:dV}
	dV(t , t^{\prime}) & \, 
	= 4 \pi R^2(t , t^{\prime}) dR 
	= 4 \pi R^2(t , t^{\prime}) \frac{\partial R}{\partial t} dt 
	= 36 \pi \, (t^{\prime})^2 \Bigl[ 1 - \Bigl( \frac{t^{\prime}}{t} \Bigr)^{1/3} \Bigr]^2 dt
\end{align}
We assume that only cuspy loops emit bursts, and the number density of cuspy loops at time $t^{\prime}$ with length between $L$ and $L+dL$ is $\nu_{\rm cuspy}(t^{\prime},L) dL$, given by \eref{eq:nu_cuspy}.  
Assuming that a burst occurs once in each loop oscillation period ($T = L / 2$), then the rate of bursts emitted from a loop of length $L$ is $\gamma_b(L) = 2 / L$.  
Since the loops are receding from the observer, the observed rate is smaller by a factor of $a(t_{\rm emitted}) / a(t_{\rm observed}) = (t_{\rm emitted} / t_{\rm observed})^{2/3}$.  
When a burst occurs, the radiation is emitted into a cone with opening angle $\theta_c(|{\bf k}|, L) \approx 0.1 / (|{\bf k}| L)^{1/3}$ with $m^{3/2} L^{1/2} < |{\bf k}| < M^{3/2} L^{1/2}$ (\cite{Long:2014mxa} and references therein).  
Then we calculate the burst rate as (see, {\it e.g.} \rref{Berezinsky:2011cp}) 
\begin{align}\label{eq:Gburst_def}
	\Gamma_{\burst}(t) dt = \int_{t_{\rm min}}^{t} dt^{\prime} \, \int_{0}^{\infty} dL \, \nu_{\rm cuspy}(t^{\prime},L) \, dV(t,t^{\prime}) \, \gamma_{b}(L) \, \left( \frac{t^{\prime}}{t} \right)^{2/3} \, \int_{\theta < \theta_{c}(L)} \frac{d\Omega}{4\pi} 
\end{align}
where we cutoff the time integral at $t_{\rm min} = t_{\rm today} / (1 + z_{\rm max})^{3/2}$.  

We can now perform the integrals in \eref{eq:Gburst_def}.  
The angular integral gives $2 \pi ( 1 - \cos \theta_c) / 4 \pi \approx \theta_c^2 / 4 \lesssim 0.0025 / (m L)$ where the inequality corresponds to $|{\bf k}| > m^{3/2} L^{1/2}$.  
To perform the loop length and time integrals, we follow the procedure of \sref{sub:Higgs_Injection} by considering separately $\nu_{\rm cuspy} = \nu_g$ and $\nu_c$, then taking the smaller of the two results.  
Using the formulae above, the remaining integrals yield 
\begin{align}\label{eq:Gburst}
	\Gamma_{\burst}(t_{\rm today}) \approx 
	10^{-3} f_c t_{eq}^{1/2} \times 
	{\rm Min} \left\{
	\begin{array}{l}
	\frac{1}{(\Gamma_g G)^{7/2} M^{7} m \, t_{\rm today}^{5/2} }
	f_1(z_{\rm max})
	\\
	\frac{m^{1/6}}{(\Gcusp)^{7/3} t_{\rm today}^{4/3} }
	f_2(z_{\rm max})
	\end{array}
	\right.
\end{align}
where $f_{1,2}(z)$ are functions of redshift that evaluate to $O(1)$ for $z_{\rm max} \approx 1$.  
If gravitational radiation is dominant then the burst rate rises rapidly with decreasing string mass scale, going like $M^{-7}$ as previously recognized by \rref{Berezinsky:2011cp}, but for lighter strings the particle radiation is dominant, and the burst rate becomes independent of $M$.  
Inserting numerical values gives $\Gamma_{\burst} < 10^{-9} \yr^{-1}$ for $\Gcusp \approx 1$, and $\Gamma_{\burst} > 1 \yr^{-1}$ for $\Gcusp < 10^{-4}$ and $M < 10^{9} \GeV$.  
Although the burst rate grows with decreasing $\Gcusp$, the amount of energy released in the form of particle radiation becomes smaller and the detection of any given burst becomes more difficult.

\section{Constraints on Model Parameters}\label{sec:Constraints}

In \sref{sec:Observables} we have endeavored to perform a model-independent analysis of the astrophysical and cosmological constraints.  
These calculations only assume that the particle radiation power can be written as in \erefs{eq:P_cusp}{eq:P_kk} for cuspy and kinky loops, respectively.  
The dimensionless coefficients, $\Gcusp$ and $\Gkk$, are treated as independent free parameters that only need satisfy $\Gcusp, \Gkk < O(1)$.  
Of all the observables surveyed in \sref{sec:Observables}, only the BBN bound actually constrains the model, as noted previously by \rref{Mota:2014uka}.  
As seen in \fref{fig:BBN_bound} the light element abundances restrict
\begin{align}\label{eq:BBN_bound_short}
	10^{-6} < \Gcusp
	\qquad {\rm for} \qquad
	10^{12} \GeV < M < 10^{15} \GeV
\end{align}
for $f_c \approx 0.5$, and the bound weakens if $f_c \ll 1$.  
In the context of specific particle radiation models, we can write the effective parameter $\Gcusp$ in terms of the Lagrangian parameters and thereby assess whether the underlying model is constrained.  

Higgs radiation results from the quadratic and linear couplings in the effective action, \eref{eq:S_eff}, and these channels correspond respectively to the expressions for $\Gcusp = \GHHcusp, \GHcusp$ given by \erefs{eq:Gamma_HH}{eq:Gamma_H}.  
In the case of the quadratic coupling we have 
\begin{align}\label{eq:GHH_too_small}
	\GHHcusp & \lesssim 10^{-7} (\gHHstr)^2 \left( \frac{m}{125 \GeV} \frac{10^{12} \GeV}{M} \right)^{1/2} 
\end{align}
with $\gHHstr = \alpha < O(1)$ given by the Higgs portal coupling.  
Over the mass range where $\Gcusp$ is constrained by BBN data, see \eref{eq:BBN_bound_short}, we see that the model predicts $\GHHcusp \ll 10^{-6}$.  
Therefore, particle radiation via the quadratic coupling alone is insufficient to constrain the model.  

Since the linear coupling violates the electroweak symmetry, the corresponding dimensionless coupling, $\GHcusp$, is proportional to the value of the Higgs condensate at the string, $\Hcond$.  
Dimensional analysis suggests that $\Hcond$ will either be set by the scale of electroweak symmetry breaking outside of the string, $\eta \approx 174 \GeV$, or by the mass scale of the string itself, $M$.  
We consider these two cases separately.  

{\bf Case 1:  Electroweak-Scale Higgs Condensate} \\
In \rref{Hyde:2013fia} we studied the structure of the dark string by numerically solving the field equations to determine the profile functions.  
Over the parameter space surveyed, we found that generally the value of the Higgs condensate at the string core remains on the order of the electroweak scale, $\Hcond \approx \eta = 174 \GeV$.  
Then $\GHcusp$ can be estimated using \eref{eq:Gamma_H} to be 
\begin{align}
	\GHcusp & \lesssim 10^{-21} (\gHstr)^2 \left( \frac{\Hcond}{174 \GeV} \frac{10^{12} \GeV}{M} \right)^2 
\end{align}
where $\gHstr$ is related to the coupling constants in the Lagrangian \cite{Hyde:2013fia}, and perturbativity requires it to not be much greater than $O(1)$.
Since $\GHHcusp \gg \GHcusp$, we see that the dominant particle radiation channel is via the quadratic interaction, as previously noted by \rref{Long:2014mxa}.  
Therefore, in the region of parameter space where $\Hcond \approx \eta$, we find that the model is entirely unconstrained, independent of the mass scale of the string.  

{\bf Case 2:  String-Scale Higgs Condensate} \\
Recently \rref{Mota:2014uka} asserted that the model under consideration here admits a region of parameter space that was overlooked by our previous analyses \cite{Hyde:2013fia, Long:2014mxa}.  
This region corresponds to $\kappa \ll \alpha < \sqrt{\kappa \lambda} < \lambda$ where $\kappa$ is the quartic self-coupling of the singlet scalar, $\lambda \approx 0.1$ is the self-coupling of the Higgs, and $\alpha$ is the Higgs portal coupling, see \eref{eq:Lint}.  
An energy minimization argument suggests that the Higgs condensate takes a value of $\Hcond \lesssim \sqrt{\alpha / \lambda} \, M$, which is not unlike the behavior in bosonic superconductivity \cite{Witten:1984eb, Haws:1988ax}.  
Using also $\gHstr \approx 10 \alpha / \kappa$ \cite{Hyde:2013fia} we can estimate 
\begin{align}
	\GHcusp \approx (10^{-2} - 10^{1}) \left( \frac{\alpha}{\sqrt{ \kappa \lambda}} \right)^4 \per
\end{align}
Then the BBN bound in \eref{eq:BBN_bound_short} implies that $(\alpha / \sqrt{\lambda \kappa}) \gtrsim 0.02$ is constrained if $10^{12} \GeV < M < 10^{15} \GeV$ and $f_c \approx 0.5$.  
The model remains unconstrained by virtue of particle emission if the string mass scale is lower $M < 10^{12} \GeV$, higher $M > 10^{15} \GeV$, or contains fewer cuspy loops $f_c \ll 0.5$.

\section{Conclusion}\label{sec:Conclusion}

Building on the work of Refs.~\cite{Hyde:2013fia, Long:2014mxa} and others, we have studied the astrophysical and cosmological constraints arising from cosmic strings in a hidden sector that couples to the SM fields through the interactions in \eref{eq:Lint}.  
In principle, the presence of relic dark strings in our universe could provide an indirect probe of the hidden sector even if the mass scale of the new physics is well above energies accessible in the laboratory.  
This is because the strings will radiate SM Higgs bosons (and also Z bosons to a lesser extent).  
In \sref{sec:Observables} we assess the impact of the Higgs decay products on nucleosynthesis, spectral distortions of the cosmic microwave background, diffuse gamma ray flux, and cosmic ray fluxes.  
BBN leads to the constraint shown in \fref{fig:BBN_bound}, but the predicted signals for the the other observables are well below the current limits.  

Since the conclusion is effectively a null result, one is inclined to ask what needs to be improved or modified if one hopes to achieve meaningful constraints.  
Increased data is not likely to change the situation.  
In some cases the data gives a measurement and in some cases it gives a bound.  
For the cases in which there is already a measurement (diffuse gamma ray and cosmic ray protons), 
it won't be possible to constrain the model with just more data.  
The predicted flux from strings is hiding under some other dominant contribution, and without being able to theoretically predict this contribution precisely, it won't matter if the precision of the data improves.  
For the cases in which there is only a bound (CMB spectral distortion 
and neutrinos), it could be possible to detect evidence of cosmic strings if the signal is eventually measured at the level predicted here.  
As we've discussed, however, the sensitivity of the near-future experiments will be insufficient to reach the level predicted from the string network except in the most optimistic parameter regime.  

Although the model considered here is found to be unconstrained by virtue of its particle emission, it is worthing noting briefly that other models of cosmic strings are known to be constrained.  
Superconducting strings emit electromagnetic radiation more copiously, and this model is constrained by CMB spectral distortions \cite{Tashiro:2012nb, Tashiro:2012nv}.  
In the case of cosmic superstrings, the reconnection probability is less than one, which enhances the loop abundance at late times, and leads to constraints from BBN and the DGRB \cite{Ookouchi:2013gwa}.  
Additionally, if the Nambu-Goto string approximation is unreliable and instead all the energy of the string network is converted into massive radiation, as argued by \rref{Hindmarsh:2008dw}, then both BBN and the DGRB lead to constraints \cite{Mota:2014uka} . 

We have identified two aspects of the calculation that have been overlooked previously.  
First, we include the contribution from both cuspy and kinky loops, whereas only one or the other has been considered in the past.  
As a result, the observables display a doubled-peaked structure when plotted against the string mass scale $M = \sqrt{\mu}$, see Figs.~\ref{fig:EvisYX},~\ref{fig:mu_dist},~and~\ref{fig:wcas}.  
The cuspy loops give the dominant contribution for large $M$, and the kinky loops dominate for small $M$.  
This behavior can be traced to the Higgs radiation powers in \erefs{eq:P_cusp}{eq:P_kk} where $\PHHcusp / \PHHkk \propto \sqrt{ML}$.  
Second, we have identified a suppression of the abundance of small loops that decay by non-gravitational radiation.  
This effect, which is discussed below \eref{eq:lr_ov_lm} (see also \aref{app:loop_dist}), arises from the non-trivial Jacobian factor that relates the loop length at different times:  if the loop decays subject to $\mu dL/dt = - P \propto L^{-\idx}$, then the Jacobian factor is proportional to $L^{\idx}$ giving a suppression for small loops.  

In conclusion, we have investigated a variety of non-gravitational signatures of dark strings in the context of the current observational status.  
We find that strings lighter than about $10^{12} \GeV$ and heavier than $10^{15} \GeV$ are not constrained by any the probes considered here.  
In this intermediate mass range, we find that BBN constrains cuspy string networks if the average power in Higgs radiation $P = \Gcusp M^2 / \sqrt{mL}$ has $\Gcusp \gtrsim 10^{-6}$; this confirms the results of \cite{Mota:2014uka}.  
This power is too large to arise from a quadratic coupling of the Higgs field to the string, as discussed below \eref{eq:GHH_too_small}, since the dimensionless coupling constant is restricted by perturbativity to be $<O(1)$.  
A large power may arise from a {\it linear coupling} of the Higgs field to the string if the Higgs condensate at the string is sufficiently large, and one obtains the bound $(\alpha / \sqrt{\lambda \kappa}) \gtrsim 0.02$ if $10^{12} \GeV < M < 10^{15} \GeV$ and the string network contains an $O(0.1)$ fraction of cuspy loops.

\acknowledgments
We are grateful to Jeffrey Hyde and Dani\`ele Steer for discussions as well as Mark Hindmarsh and Eray Sabancilar for comments on an early version of this draft.
This work was supported by the Department of Energy at ASU.  

\begin{appendix}

\section{Derivation of Loop Length Distribution}\label{app:loop_dist}

We derive the loop length distribution following \rref{Blanco-Pillado:2013qja}, which considers loop decay via gravity wave emission.  
We extend their calculation to include loops that decay by particle emission.  

Let $f(t,L,p) dt \, dL \, dp$ be the number of loops produced per comoving volume between time $t$ and $t+dt$ with length between $L$ and $L+dL$ and momentum between $p$ and $p+dp$.  
Let $n(t,L,p) dL \, dp$ be the number of loops per comoving volume at time $t$ with length between $L$ and $L+dL$ and momentum between $p$ and $p+dp$.  
Then we have 
\begin{align}
	n(t,L,p) & = \int_{0}^{t} \,dt_{i}  f(t_{i}, L_{i}, p_{i}) \frac{\partial L_{i}}{\partial L} \frac{\partial p_{i}}{\partial p}
\end{align}
where $L_{i}$ is the length of a loop at time $t_{i}$ that will later have length $L$ at time $t>t_{i}$, and similarly $p_{i}$ is the momentum of a loop at time $t_{i}$ that will later have momentum $p$ at time $t>t_{i}$.  
Simulations of string networks tell us the loop production function in the absence of loop decay, and the decay is taken into account by the transformation from $L$ to $L_{i}$ and $p$ to $p_{i}$.

The momentum is assumed to evolve solely due to cosmological redshift:  
\begin{align}
	p_{i} = p \frac{a}{a_{i}}
	\qquad , \qquad
	\frac{\partial p_{i}}{\partial p} = \frac{a}{a_{i}} \geq 1
\end{align}
where $a_{i}$ is the FRW scale factor at time $t_{i}$ that later becomes $a$ at time $t$.  
In the radiation- and matter-dominated eras, we have $a_{i} / a  = t_{i}^{\nu} / t^{\nu}$ with $\nu = 1/2$ and $\nu = 2/3$, respectively.  
The particle horizon at time $t$ is given by $d_h(t) = a \int_{0}^{t} dt_{i} / a_{i} = t / (1-\nu)$.  

It is convenient to move to scaling coordinates
\begin{align}\label{eq:scaling_coords}
	\alpha = \frac{L}{d_h(t)}
	\qquad , \qquad
	\alpha_{i} = \frac{L_{i}}{d_h(t_i)} \per
\end{align}
Let $f(\alpha,p)d\alpha dp$ be the number of loops formed in a {\it physical} volume $d_h^3$ in a time $d_h$ with $\alpha$ and $p$ in the ranges $d\alpha$ and $dp$, and let $n(t,\alpha,p)d\alpha dp$ be the number of loops present in a {\it physical} volume $d_h^3$ at a time $t$ with $\alpha$ and $p$ in the ranges $d\alpha$ and $dp$.  
If the system is in the scaling regime, then $n$ only depends on time through the scaling length $\alpha$, but in general (and specifically, in the case of particle radiation that we consider here) $n$ will be a function of both $t$ and $\alpha$ separately.  
Then, we have the relationships\footnote{
That is, $f(\alpha,p) = \int_{d_h} dt \int_{d_h^3} d^3 x \, \frac{f(t,L,p)}{a^3} \frac{\partial L}{\partial \alpha}$.  
}
\begin{align}\label{eq:scaling}
	f(\alpha,p) &= \frac{d_h^5}{a^3} f(t,L,p) \com \nn
	n(t,\alpha,p) &= \frac{d_h^4}{a^3} n(t,L,p) \com
\end{align}
and
\begin{align}
	n(t,\alpha,p) 
	= (1-\nu) \int_{\infty}^{\alpha} d\alpha_{i} \frac{t^{4-2\nu}}{(t_{i})^{5-2\nu}} f(\alpha_{i}, p \frac{t^{\nu}}{(t_{i})^{\nu}}) \frac{\partial L_{i}}{\partial L} \frac{\partial t_{i}}{\partial \alpha_{i}} 
\end{align}
where the quantities with an $i$ subscript are understood to be functions of $\alpha_{i}$.  
Since we are not interested in the distribution of momenta, we integrate $f(\alpha) = \int_{0}^{\infty} dp \, f(\alpha,p)$ and $n(t,\alpha) = \int_{0}^{\infty} dp \, n(t,\alpha,p) $.  
This gives the ``master formula'':
\begin{align}
	n(t,\alpha) = (1-\nu) \int_{\infty}^{\alpha} d\alpha_{i} \, \frac{t^{4-3\nu}}{(t_{i})^{5-3\nu}} f(\alpha_{i}) \frac{\partial L_{i}}{\partial L} \frac{\partial t_{i}}{\partial \alpha_{i}} \per
\end{align}
To proceed further, we must specify the loop decay process (this gives a relationship between $t_{i}$ and $\alpha_{i}$) and the cosmological epoch (this gives $\nu$ and $f(\alpha_{i})$).  

As we discuss in \sref{sub:LoopDecay}, the evolution of a given loop is determined by solving the initial value problem 
\begin{align}\label{eq:dLdt_appendix}
	M^2 \frac{dL}{dt} = - P(L)
	\qquad , \qquad
	L(t=t_{i}) = L_{i} 
\end{align}
where $P(L)$ is the rate at which a loop of length $L$ at time $t$ loses energy into gravitational radiation and particle emission.  
We assume that the power can be parametrized as 
\begin{align}
	P(L) = \pp \frac{M^2}{(mL)^{\idx}} 
\end{align}
where $\pp$ is a dimensionless parameter. 
For gravitational radiation $\idx = 0$ and $\pp = \Gamma_g G M^2$, and for all the particle radiation channels discussed in \sref{sec:SM_Particle_Prod} $\idx \neq 0$ and $\pp < 1$ is a dimensionless coefficient.  
The solution of \eref{eq:dLdt_appendix} is  
\begin{align}\label{eq:dLdL}
	(L_{i})^{\idx+1} & = L^{\idx +1} + (1+\idx) (t - t_{i}) \frac{\pp}{m^{\idx}} 
	\qquad , \qquad 
	\frac{\partial L_{i}}{\partial L} = \frac{L^{\idx}}{L_{i}^{\idx}} = \frac{\alpha^{\idx} t^{\idx}}{\alpha_{i}^{\idx} t_{i}^{\idx}} \leq 1
	\per
\end{align}
In terms of the scaling coordinates, the solution is 
\begin{align}\label{eq:tp_to_ap}
	(\alpha_{i})^{\idx+1} (t_{i})^{\idx+1} + (1+\idx)(1-\nu)^{\idx+1} \frac{\pp}{m^{\idx}} t_{i} = \alpha^{\idx +1} t^{\idx+1} + (1+\idx)(1-\nu)^{\idx+1} \frac{\pp}{m^{\idx}} t 
\end{align}
from which we obtain
\begin{align}\label{eq:tp_soln}
	\frac{\partial t_{i}}{\partial \alpha_{i}} = - \frac{t_{i}}{\alpha_{i} + (1-\nu)^{\idx+1} \frac{\pp}{(m \alpha_{i} t_{i})^{\idx}}} \per
\end{align}
Using $\partial L_{i} / \partial L$ and $\partial t_{i} / \partial \alpha_{i}$ from above, the master formula becomes
\begin{align}\label{eq:master_2}
	n(t,\alpha) = (1-\nu) \int_{\alpha}^{\infty} d \alpha_{i} \left( \frac{t}{t_{i}} \right)^{4-3\nu+\idx} \left( \frac{\alpha}{\alpha_{i}} \right)^{\idx} \frac{f(\alpha_{i})}{\alpha_{i} + (1-\nu)^{\idx+1} \frac{\pp}{(m \alpha_{i} t_{i})^{\idx}}}
\end{align}
where $t_{i}$ is expressed as a function of $\alpha_{i}$, $\alpha$, and $t$ by solving \eref{eq:tp_to_ap}.  

It is not always possible to obtain the solution of \eref{eq:tp_to_ap} in closed form.  
However, it turns out that the linear term on the left hand side is always negligible,\footnote{
The ratio of the first to second term is $O(\alpha_{i}^{\idx +1} (mt_{i})^{\idx} / \pp)$.  
The factor $(mt_{i})^{\idx}$ is large for cosmological time scales, and although the factor $\alpha_{i}^{\idx+1}$ could be arbitrarily small in principle, we will see in the paragraph below that the integral over $\alpha_{i}$ is dominated by $\alpha_{i} = O(0.01)$.  
For the case of gravitational radiation with $\idx=0$ we have $\pp = O(G M^2) \ll 1$.  
}
and the solution is simply 
\begin{align}\label{eq:tsol_r}
	t_{i} = 
	\frac{1}{\alpha_{i}} \Bigl[
	\alpha^{\idx +1} t^{\idx+1} + (1+\idx)(1-\nu)^{\idx+1} \frac{\pp}{m^{\idx}} t 
	\Bigr]^{1/(\idx+1)}
	\per
\end{align}
The same approximation allows us to neglect the second term in the denominator of \eref{eq:master_2}, since it arose from the linear term in \eref{eq:tp_to_ap}.  
Inserting \eref{eq:tsol_r} into the master formula gives
\begin{align}\label{eq:master_3}
	n(t,\alpha) = 
	 \frac{(1-\nu) \, t^{4-3\nu+\idx} \, \alpha^{\idx}}{[\alpha^{\idx +1} t^{\idx+1} + (1+\idx)(1-\nu)^{\idx+1} \frac{\pp}{m^{\idx}} t ]^{\frac{4-3\nu+\idx}{\idx+1}}} 
	\int_{\alpha}^{\infty} d \alpha_{i} 
	& (\alpha_{i})^{3-3\nu} f(\alpha_{i})
	\per 
\end{align}

The loop formation function $f(\alpha_{i})$ is extracted from simulations of string networks.  
During the radiation-dominated era ($\nu=1/2$), the formation function is well-approximated by a Dirac delta function \cite{Blanco-Pillado:2013qja}
\begin{align}\label{eq:fr}
	f_r(\alpha_{i}) \approx \frac{\zeta_{r}}{\alpha_{i}^{3/2}} \delta(\beta_{r} - \alpha_{i})
\end{align}
where $\zeta_{r} = 1.04$ and $\beta_{r} = 0.05$.  
Inserting $f_r$ into \eref{eq:master_3} gives
\begin{align}\label{eq:nr_approx}
	n_r(t, \alpha) = \frac{\zeta_{r}}{2} \, \Theta(\beta_{r}-\alpha) 
	\frac{t^{5/2+\idx} \alpha^{\idx} }{[\alpha^{\idx +1} t^{\idx+1} + \frac{(1+\idx)}{2^{\idx+1}} \frac{\pp}{m^{\idx}} t ]^{\frac{5/2+\idx}{\idx+1}}} 
	\per
\end{align}
During the matter-dominated era ($\nu=2/3$) the formation function instead takes the form \cite{Blanco-Pillado:2013qja}
\begin{align}\label{eq:fm}
	f_m(\alpha_{i}) \approx \frac{\zeta_{m}}{\alpha_{i}^{1.69}} \Theta(\beta_{m} - \alpha_{i})
\end{align}
where $\zeta_{m} = 5.34$ and $\beta_{m} = 0.06$.  
Inserting $f_m$ into \eref{eq:master_3} gives
\begin{align}\label{eq:nm_approx}
	n_m(t, \alpha) = \frac{\zeta_{m}}{3} 
	\frac{\beta_{m}^{0.31} - \alpha^{0.31} }{0.31} 
	\frac{t^{2+\idx} \alpha^{\idx} }{[\alpha^{\idx +1} t^{\idx+1} + \frac{(1+\idx)}{3^{\idx+1}} \frac{\pp}{m^{\idx}} t ]^{\frac{2+\idx}{\idx+1}}} 
\per
\end{align}

Radiation-matter equality occurs at a time $t = t_{eq}$.  
A loop with scaling length $\alpha$ at time $t$ has a length $L_{eq}$ at time $t_{eq}$, which is found by solving \eref{eq:dLdL} with $L = 3 t \alpha$.  
The corresponding scaling length is $\alpha_{eq} = L_{eq} / 2 t_{eq}$ and hence 
\begin{align}\label{eq:daeqda}
	\alpha_{eq}^{\idx+1} = \left( \frac{3t}{2t_{eq}} \right)^{\idx+1} \alpha^{\idx+1} + \frac{ (1+\idx) (t-t_{eq}) \pp }{ (2 t_{eq})^{\idx+1} m^{\idx}} 
	\qquad , \qquad
	\frac{\partial \alpha_{eq}}{\partial \alpha} = \left( \frac{3t}{2t_{eq}} \right)^{\idx+1} \frac{\alpha^{\idx}}{\alpha_{eq}^{\idx}} \per
\end{align}
In the matter era, the population of radiation era relic loops is given an appropriate transformation of the distribution at radiation-matter equality:\footnote{The number of loops per horizon volume at time $t_{eq}$ with scaling length between $\alpha_{eq}$ and $\alpha_{eq} + d\alpha_{eq}$ is $n_r(t_{eq},\alpha_{eq}) d\alpha_{eq}$. Multiplying by $(a_{eq} / 2 t_{eq})^3$ gives the number per comoving volume, and multiplying by $(3 t / a)^3$ gives the number per physical horizon volume at a later time $t$.  The final Jacobian factor $\partial \alpha_{eq} / \partial \alpha$ transforms the differential element to give $n_r(t > t_{eq}, \alpha) d\alpha$.  }
\begin{align}
	n_r(t>t_{eq},\alpha) = n_r(t_{eq}, \alpha_{eq}) \frac{a_{eq}^3}{a^3} \frac{(3t)^3}{(2t_{eq})^3} \frac{\partial \alpha_{eq}}{\partial \alpha} \per
\end{align}
Using $a \sim t^{2/3}$ during the matter era and inserting the solution from \eref{eq:nr_approx} gives
\begin{align}\label{eq:nr_relic}
	n_r(t>t_{eq},\alpha) = 
	\frac{\zeta_{r}}{2^{5/2}} \, \Theta(\beta_{r}-\alpha_{eq}(t)) 
	\frac{3^{3/2} t^{2+\idx} t_{eq}^{1/2} \alpha^{\idx} }{[\alpha^{\idx +1} t^{\idx+1} + \frac{(1+\idx)}{3^{\idx+1}} \frac{\pp}{m^{\idx}} t ]^{\frac{5/2+\idx}{\idx+1}}} 
\end{align}
where $\alpha_{eq}(t)$ is given by \eref{eq:daeqda}.  

We convert back to physical coordinates using \eref{eq:scaling}, $\nu(t,L) = n(t,L) / a^3 = n(t,\alpha) / d_h^4$, such that $\nu(t,L) dL = n(t,L) dL / a^3$ is the number density of loops per physical volume at time $t$ with length between $L$ and $L+dL$.  
Applying these transformations to Eqs.~(\ref{eq:nr_approx}),~(\ref{eq:nm_approx}),~and~(\ref{eq:nr_relic}) gives the length distribution of loops during the radiation era, the length distribution of loops formed during the matter era, and the length distribution loops surviving from the radiation era into the matter era
\begin{subequations}\label{eq:n_physical}
\begin{align}
	\nu_{r}(t<t_{eq},L) & = 
	\frac{\zeta_{r}}{2^{5/2}} \, 
	\Theta(2 t \beta_{r}-L) \, 
	\frac{1}{t^{3/2}} \, 
	\frac{L^{\idx} }{L_0(t,L)^{5/2+\idx}}
	\\
	\nu_{m}(t>t_{eq},L) & = 
	\frac{\zeta_m}{3^{3}} \, 
	\frac{\beta_{m}^{0.31} - (L/3t)^{0.31} }{0.31} \, 
	\frac{1}{t^2} \, 
	\frac{L^{\idx} }{L_0(t,L)^{2+\idx} } 
	\\
	\nu_{r}(t>t_{eq},L) & = 
	\frac{\zeta_{r}}{2^{5/2}} \, 
	\Theta(2 t_{eq} \beta_{r} - L_{eq}(t,L)) \, 
	\frac{t_{eq}^{1/2}}{t^2} \, 
	\frac{L^{\idx}}{ L_0(t,L)^{5/2 + \idx}} 
\end{align}
\end{subequations}
where
\begin{align}
	L_{eq}(t,L) & = \bigl[ L^{\idx+1} + (1+\idx)(t-t_{eq}) \frac{\pp}{m^{\idx}} \bigr]^{1/(\idx+1)} \\
	L_0(t,L) & = \bigl[ L^{\idx+1} + (1 + \idx) \frac{\pp}{m^{\idx}} t \bigr]^{1 / (\idx +1)}
\end{align}
are the length of a loop at radiation-matter equality and at $t=0$, respectively, that later has length $L < L_{eq}, L_0$ at time $t$.  

Note especially the factors of $(L / L_0)^{\idx}$ in \eref{eq:n_physical}.  
These factors arose from the Jacobian in \erefs{eq:dLdL}{eq:daeqda} associated with the non-linear coordinate transformation between $L(t)$ and $L_{i}(t_{i})$.

\end{appendix}

\bibliographystyle{h-physrev5}
\bibliography{refs--Dark_Strings}

\end{document}